\documentclass[10pt]{article}
\usepackage{amsmath}
\usepackage{amssymb}
\usepackage{graphicx}
\usepackage{cite}
\usepackage{color} 

\usepackage{setspace} 

\usepackage[labelfont=bf,labelsep=period,justification=raggedright]{caption}

\makeatletter
\renewcommand{\@biblabel}[1]{\quad#1.}
\makeatother

\date{}

\newcommand{\bb}{\begin{eqnarray}}
\newcommand{\ee}{\end{eqnarray}}

\begin{document}

\begin{flushleft}
{\Large
\textbf{Contact patterns among high school students}
}
\\
Julie Fournet$^{1}$, 
Alain Barrat$^{1,2,\ast}$
\\
\bf{1} Aix-Marseille Universit\'e, Universit\'e de Toulon, CNRS, CPT UMR 7332, 13288 Marseille, France
\\
\bf{2} Data Science Laboratory, Institute for Scientific Interchange (ISI) Foundation, Torino, Italy
\\
$\ast$ E-mail: alain.barrat@cpt.univ-mrs.fr
\end{flushleft}

\section*{Abstract}

Face-to-face contacts between individuals contribute to shape social networks and play an important role in determining how infectious
diseases can spread within a population. It is thus important to obtain accurate and reliable descriptions of human contact patterns
occurring in various day-to-day life contexts. Recent technological advances and the development of wearable sensors
able to sense proximity patterns have made it possible to gather data giving access to time-varying contact networks of individuals in 
specific environments. 
Here we present and analyze two such data sets describing with high temporal resolution the contact patterns of students in a high school. We
define contact matrices describing the contact patterns between students of different classes and 
show the importance of the class structure. We take advantage of the 
fact that the two data sets were collected in the same setting during several days in two successive years to perform a longitudinal
analysis on two very different timescales. We show the high stability of the contact patterns across days and across years: the statistical distributions
of numbers and durations of contacts are the same in different periods, and we observe
a very high similarity of the contact matrices measured in different days or different years. The rate of change of the contacts of each individual from one day to
the next is also similar in different years. 
We discuss the interest of the present analysis and data sets for various fields, including in social sciences in order to better understand and model
human behavior and interactions in different contexts, and in epidemiology in order to inform models describing the spread of infectious diseases and
design targeted containment strategies.

\section*{Introduction}

Reliable detailed information on the contact patterns occurring between individuals in day-to-day life contexts carries a great value 
in fields such as social sciences or epidemiology of infectious diseases, in which human interactions are of primary importance.
In particular, the investigation of contact patterns
in contexts where a substantial amount of social mixing between individuals is expected, such as schools, high schools
or workplaces, represents an important research goal. 
For instance, the strong mixing of school children favors the spread of
infectious diseases in school environmentx and makes them an important source of infection into households from 
where infections can spread further~\cite{Longini:1982,Viboud:2004}. In such contexts, 
a precise description of human contacts can help identify possible contagion pathways,  design realistic models of epidemic spread and 
design and evaluate containment strategies such as  targeted vaccination, social distancing or school or workplace closures.

Many efforts have therefore been devoted to the collection of data on human contact patterns in various settings and environments 
in the last years~\cite{Read:2012}. The research community has moreover started to systematically take advantage of recent
technological advances to move from methods ranging from diaries and surveys~\cite{Edmunds:1997,Mossong:2008,Read:2008,Zagheni:2008,Mikolajczyk:2008,Conlan:2011,Smieszek:2012,Potter:2012,Danon:2013,Smieszek:2014}
to new technologies based on wearable sensors 
able to detect close proximity~\cite{Pentland:2008,Salathe:2010,Hornbeck:2012,Smieszek:2014,Sekara:2014,Barclay:2014}
and even face-to-face contacts of  individuals~\cite{SocioPatterns,Cattuto:2010,Isella:2011a,Stehle:2011a,Stehle:2011b,Isella:2011b,Vanhems:2013}.
Biases due to self-reporting are thus avoided \cite{Smieszek:2012,Smieszek:2014} and high-resolution data can be collected
in an objective way, allowing to parametrize and inform data-driven models describing human behavior \cite{Stehle:2010,Zhao:2011,Starnini:2013}
and epidemic spread in specific settings \cite{Stehle:2011a}.

Thanks to the use of wearable sensors, face-to-face
contact data collected in unsupervised fashion in different contexts
have thus started to become available, providing the beginning of an ``atlas'' of human contacts: the contexts
investigated include conferences 
\cite{Cattuto:2010,Isella:2011b,Stehle:2011a}, a museum \cite{Isella:2011b}, a primary school
\cite{Stehle:2011b}, in which a strong impact of the class and age structure on the contact patterns was evidentiated,
and hospitals \cite{Isella:2011a,Vanhems:2013}. Many contexts remain however to be investigated.
Moreover, the longitudinal dimension of contact patterns has barely been studied \cite{Vanhems:2013}, and data sets describing face-to-face
contacts in a given population or in a given setting but at different times have to our knowledge not been collected or analyzed.
Here, we 
partially address this issue, and give a new contribution to the collection of contact patterns in diverse environments,
by presenting two high-resolution data sets describing the contacts between students in a high school.
These data were collected in a French high school in 2011 and 2012 using a proximity-sensing platform based on wearable sensors. 
We investigate the mixing patterns of students as described by their high-resolution temporal contact network. We study how the mixing is driven
by the repartition of students into classes and investigate if gender differences have an impact on contact patterns, as observed
in primary school \cite{Stehle:2013}.
We moreover study the evolution of the contact patterns on two widely distinct different timescales: on the one hand, we examine the stability of the contact
networks and mixing patterns in a given population of students from one day to the next; on the other hand,  we take advantage
of the fact that the two data sets were collected in the same environment (even if the students changed from one year to the next) 
to study the  long term stability of contact patterns in the high school. 
Finally, we 
make available on a dedicated webpage 
(\verb+http://www.sociopatterns.org/datasets/+)
a novel instance of a high-resolution time-varying network of contacts between individuals, which can
be of interest to the wide interdisciplinary research community studying complex and temporal networks.


\section*{Methods}

\subsection*{Study design, high school context, and data collection}

We collected data on the close proximity face-to-face encounters of high school students of several classes in 
Lyc\'ee Thiers, Marseilles, France during 4 days (Tuesday to Friday) in Dec. 2011 and during
7 days (from a Monday to the Tuesday of the following week) in Nov. 2012. The data collection involved
students of three different classes in 2011 (gathering 118 students) and of five classes (gathering 180 students) 
in 2012. The three classes participating in 2011 were among the five of 2012, albeit with different students.
These classes, called ``classes pr\'eparatoires'', are specific to the
French schooling system. They gather students for studies that take place after the end of the usual high school studies and 
last two years. During these two years, students still study in a high school environment but are de facto mostly separated from the
younger high school students, { as their classes are located in a different part of the high school building and as they take their lunches separately}. 
At the end of these two years, students go through competitive exams yielding admission to various
higher education colleges. 

The data collection involved classes corresponding to the second year of such studies, in which
students focus on different topics: ``MP'' classes focus more on mathematics and physics, ``PC'' classes on physics and chemistry,
and ``PSI'' classes on engineering studies. All students in this second year must prepare a small project that they present at the
exam, and several students could build a project based on their
participation to the data collection, together with the use of the collected data in some small scale analysis or numerical
simulation. Thanks to their involvement, the participation of students to the data collection was close to $100\%$.
{ We however limited the data collection to those classes in which at least one student would prepare a project
based on the data, as engagement is a well-known issue that is efficiently solved through this kind of incentive \cite{Conlan:2011}. 
In particular, first year students did not participate, and some classes participating in the 2012 study were not included in the first
study (2011). The comparison between the data sets of the two
distinct years will be performed taking this point into account.
Overall, we thus obtain data concerning the contacts of a specific set of students for each year, corresponding to
a specific set of classes, and the data 
do not contain contacts with individuals not taking part to the study. Importantly, the study considers whole classes, and not a subset of students
in each class (this is important because most contacts occur within classes as we will see later on). Moreover, second year students themselves
assert that they almost do not have any contacts with first year students.}

Data were gathered using the measurement platform developed by the SocioPatterns collaboration \cite{SocioPatterns}: it is based on 
wearable sensors that are embedded in unobtrusive wearable badges and exchange ultra-low power radio packets in order
to detect close proximity of individuals wearing them \cite{SocioPatterns, Cattuto:2010}. As described in detail elsewhere 
\cite{Cattuto:2010,Stehle:2011b,Isella:2011a,Vanhems:2013}, 
the power level is tuned so that devices can exchange packets only when located within 1-1.5 meters of one another,
and individuals are asked to wear the devices on their chests using lanyards, ensuring that the devices of two individuals 
can only exchange radio packets when the persons are facing each other.
Moreover, the wearable sensors are tuned so that the face-to-face proximity of two individuals wearing them can be assessed over an interval of $20$ seconds with a probability in excess of $99\%$. Two individuals are thus said to be in contact if their badges exchange radio packets during a 20-second time window, and the contact
event is considered interrupted if the badges do not exchange packets over a 20-second interval. 
Finally, the information on face-to-face proximity events detected by the wearable sensors is relayed to radio receivers installed throughout the high school:
contacts occurring outside the school premises were not measured.

\subsection*{Ethics and privacy}

Before the study, students and teachers were informed on the details and aims of the study. 
A signed informed consent was obtained for each participating individual (no minors were involved
as students of these classes were all aged at least 18).
All participants were given a wearable sensor and asked to wear it at all times
in the high school.  No personal information was collected: the only information 
associated with the unique identifier of each badge was the class 
and the gender of the corresponding individual.
The ethics committee responsible for this kind
of data collection, which is the French national bodies responsible for ethics and privacy, 
namely
the Commission Nationale de l'Informatique et des 
Libert\'es (CNIL, http://www.cnil.fr) 
approved the study, as well as the high school authorities.\\

\subsection*{Data analysis}

At the most detailed level, the data collecting infrastructure yields a temporal network of contacts with a temporal resolution
of $20$ seconds \cite{Cattuto:2010}. 
Starting from these data, we analyze the 
patterns of contacts between students and between classes at different temporal and structural aggregation levels. 
In the following, $i$ and $j$ denote individual sensor identification numbers, while $X$ and $Y$ denote classes. 

We first compute the number of contact events of each individual, the statistical distribution of the duration of such events,
and of the time between successive contacts of an individual. We moreover build aggregated contact networks on several
time windows: in each of these networks, nodes represent individuals and a weighted link between two nodes 
represents the fact that the two corresponding individuals have been in contact at least once during the aggregation time window. 
For each time window, we can define the following quantities:
\begin{itemize}
\item $e_{ij}$ is 1 if and only if $i$ and $j$ have been in contact: this corresponds to the adjacency matrix of the contact network
aggregated over the considered time window,

\item the degree $k_i$ of a node gives the number of distinct persons with whom $i$ has been in contact during the time window,

\item $n_{ij}$ gives the number of contact events recorded between $i$ and $j$ ($n_{ij}=n_{ji}$) during the time window,

\item the weight $w_{ij}$ of the link between $i$ and $j$ is defined by the 
cumulative duration of the $n_{ij}$ contacts between $i$ and $j$ ($w_{ij}=w_{ji}$) which occurred during the time window,

\item the strength of a node $i$, $s_i = \sum_{j} w_{ij}$, gives the sum of the durations of the contacts of individual $i$ during the time window.
\end{itemize}

We investigate the statistical distributions of the degrees and weights in the aggregated networks. For each distribution, we compute
average and coefficient of variation squared $CV^2$ (squared ratio of the standard deviation to the mean of the distribution: $CV^2<1$ 
corresponds to distributions with low variance, while $CV^2>1$ is obtained for high-variance distributions). 

We also compare the properties of networks aggregated
over different time windows, and the evolution of some nodes' properties (degree and strength)
when the aggregation time window length increases. At a finer resolution, 
we moreover investigate the similarly between the neighborhoods of a given node in contact networks aggregated over different
periods. For instance, for daily aggregated networks,  the similarity between the neighborhoods of an individual $i$ in the contact 
networks measured in two different days denoted 1 and 2 is measured through the cosine similarity 
$$
\sigma^{1,2}(i)=\frac{\sum_j w_{ij,1}w_{ij,2}}{\sqrt{\sum_j {w}_{ij,1}^2}\sqrt{\sum_j w_{ij,2}^2}},
$$
where $w_{ij,d}$ is the weight of the link between $i$ and $j$ in the contact network of day $d$, i.e., the cumulative duration of the contacts between $i$ and $j$
occurring on day $d$. The cosine similarity takes values between $0$ and $1$: it is equal to $0$ if $i$ had contact with strictly different
individuals in the two days considered, and to $1$ if $i$ had contacts with the same persons in both days,
with proportional durations.

As students are divided into classes, we moreover aggregated the data in order to study the mixing patterns between classes.
The number of students in class $X$ is denoted by $n_X$. We consider the following quantities, aggregated over each 
time window of interest:
\begin{itemize}

\item the number of edges between students of class $X$ with students of class $Y$  in the aggregated
contact network: $E_{XY}=\sum_{i \in X, j \in Y}e_{ij}$ for $X \neq Y$ (and $E_{XX}=\frac{1}{2}\sum_{i, j \in X}e_{ij}$),

\item the density of edges between  class $X$ and class $Y$: $\rho_{XY} = E_{XY}/ E^{max}_{XY}$, where 
$E^{max}_{XY} = n_{X} n_{Y}$  is the maximum possible number of edges between class $X$ and class $Y$
($E^{max}_{XX} =  n_{X}(n_{X}-1)/2$).

\item the total number of contacts between students of class $X$ with students of class $Y$ : 
$N_{XY}=\sum_{i \in X, j \in Y}n_{ij}$ (for $X=Y$ we have $N_{XX}=\frac{1}{2}\sum_{i, j \in X}n_{ij}$),

\item the average number of contacts of a student of class $X$ with students of class $Y$ $n_{XY} = \frac{N_{XY}}{n_X}$,

\item the total time spent in contact between students of class $X$ with students of class $Y$ : $W_{XY}=\sum_{i \in X, j \in Y}w_{ij}$ 
(for $X=Y$ we have $W_{XX}=\frac{1}{2}\sum_{i, j \in X}w_{ij}$),

\item the average time spent by a student of class $X$ in contact with students of class $Y$ : $w_{XY}= \frac{W_{XY}}{n_X}$.
\end{itemize}

The quantities $W_{XY}$, $E_{XY}$ and $\rho_{XY}$ define contact matrices that describe the mixing patterns between 
the classes. In order to investigate the temporal stability of these patterns, we moreover consider 
the similarity between two matrices $A$ and $B$ of size $n\times n$, describing the contact matrices in different
time windows, defined as
$$
\sigma_{A,B}=\frac{\sum_{i,j=1}^{n} A_{ij}B_{ij}}{\sqrt{\sum_{i,j=1}^{n}A_{ij}^2}\sqrt{\sum_{i,j=1}^{n}B_{ij}^2}}.
$$
{ This similarity takes values between $0$ and $1$: it is equal to $1$ if the elements of both matrices are proportional
($A_{ij} = c B_{ij}$ for all $i,j$, with $c$ independent of $i$ and $j$), and $0$ if each time one element of $A$ is non-zero, the corresponding
element of $B$ is zero and vice-versa.}


\section*{Results}

\subsection*{Global analysis}

The class names, the number  of individuals of each gender and in each class are given respectively in Tables \ref{table:1} and \ref{table:2} for each year.
In order to avoid repetitions, we mainly report here results corresponding to the data collection performed in 2012. A comparison between
the two data sets is performed later on.

\begin{table}[!h]
\begin{tabular}{|l||c|c|c|}
\hline
Class name &  Number of individuals & Male & Female\\
\hline
PC&31&16&15\\
PC$^*$&45&32&13\\
PSI$^*$&42&32&10\\
\hline
Total&118 & 80& 38 \\
\hline
\end{tabular}
\caption{Classes involved in the 2011 data collection}
\label{table:1}
\end{table}

\begin{table}[!h]
\begin{tabular}{|l||c|c|c|c|c|c|c|}
\hline
Class name &  Number of individuals & Male & Female \\
\hline
MP$^*$1&31&27&4 \\
MP$^*$2&35&27&8 \\
PC&38&24&14 \\
PC$^*$&35&26&9 \\
PSI$^*$&41&29&12\\
\hline
Total& 180 & 133& 47 \\
\hline
\end{tabular}
\caption{Classes involved in the 2012 data collection}
\label{table:2}
\end{table}

During the 7 days of data collection of 2012, 19,774 contact events were registered, corresponding to a cumulative duration of 900,940s (approx. 250 hours). 
Table \ref{table:3} reports the number of durations of contacts registered in each day.

\begin{table}
\begin{tabular}{|l||l|l|l|l|}
\hline
&  Number of contacts  & \multicolumn{3}{c|}{Cumulative duration of contacts}\\
\hline
Day &Number (\% of total) &Seconds (\% of total) & Minutes &Hours\\
\hline
1st Monday&4,191 (21.2)&199,140 (22.1)&3,319&55\\
1st Tuesday&3,170 (16)&132,720 (14.7)&2,212&37\\
Wednesday&1,547 (7.8)&64,540 (6.4)&965&16\\
Thursday&2,641 (13.4)&106,920 (11.9)&1,782&30\\
Friday&3,184 (16.1)&154,360 (17.1)&2,573&43\\
2nd Monday&2,988 (15.1)&156,360 (17.4)&2,606&43\\
2nd Tuesday&2,053 (10.4)&93,540 (10.4)&1,559&26\\
\hline
Total&19,774&900,940&15,016&250\\
\hline
\end{tabular}
\caption{Number and duration of contacts in the different days of the 2012 data collection that lasted 7 days (one full week plus monday and tuesday
of the following week).}
\label{table:3}
\end{table}

\begin{figure}
\includegraphics[width=.8\textwidth]{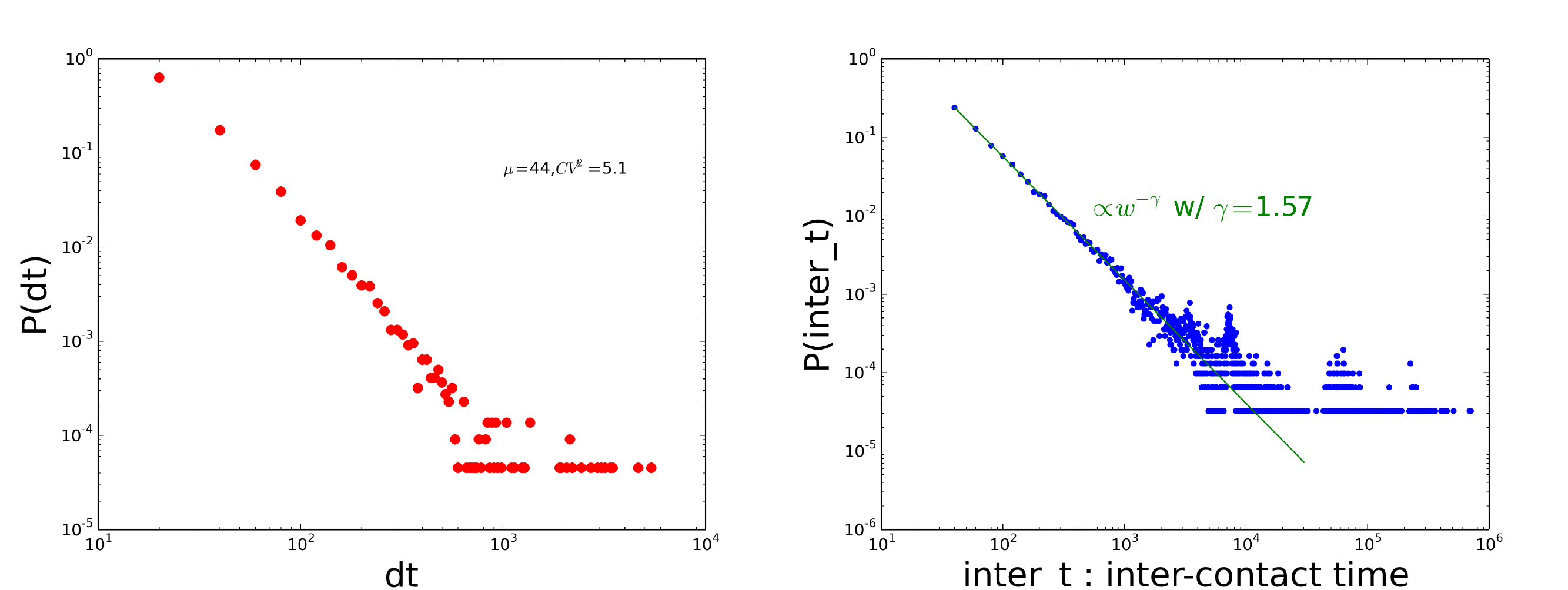}
\caption{Left: Distribution $P(dt)$ of contact durations: probability for a contact to last $dt$,
Right: Distribution $P(inter\_t)$ of inter-contact durations: probability that the time elapsed between two successive contacts is $inter\_t$.}
\label{fig:pdt} 
\end{figure}

Figure \ref{fig:pdt} reports the distributions of contact and inter-contact durations measured over the whole data collection. 
Most contacts are short, but contacts of very different durations are observed, including long ones. 
While the average duration of a contact is 44 seconds, and 88\% of the contacts last less than 1 minute, more than 1\% of the contacts last at least 5 minutes. 
The strong variability in contact durations is shown by the large value of the
squared coefficient of variation of the distribution, $CV^2=5.1$. In fact, the distribution is heavy-tailed:
as already observed in previous studies measuring the durations of contact events between individuals 
(see e.g., \cite{Cattuto:2010,Salathe:2010}), and no characteristic contact time scale can be defined.
{ In other terms, the average contact duration is not representative and both much shorter and much longer contacts 
can be observed with non-negligible probabilities.
For transmissible diseases for which the transmission probability between two individuals 
depends on their time in contact, this means that different contacts might yield very different 
transmission probabilities: many contacts are very short and correspond to a small transmission probability, 
but some are much longer than others, and could therefore play a crucial role in disease dynamics. }
The inter-contact durations is as well very broad (close to a power-law with exponent smaller than 2): most intervals between 
successive contacts are very short,
but very long durations are also observed. This highlights the burstiness of human contacts, a well known feature of human dynamics  observed
in a variety of systems  driven by human actions~\cite{Barabasi:2005,Barabasi:2010}.

\subsection*{Contact matrices}

\begin{figure}
\includegraphics[width=.8\textwidth]{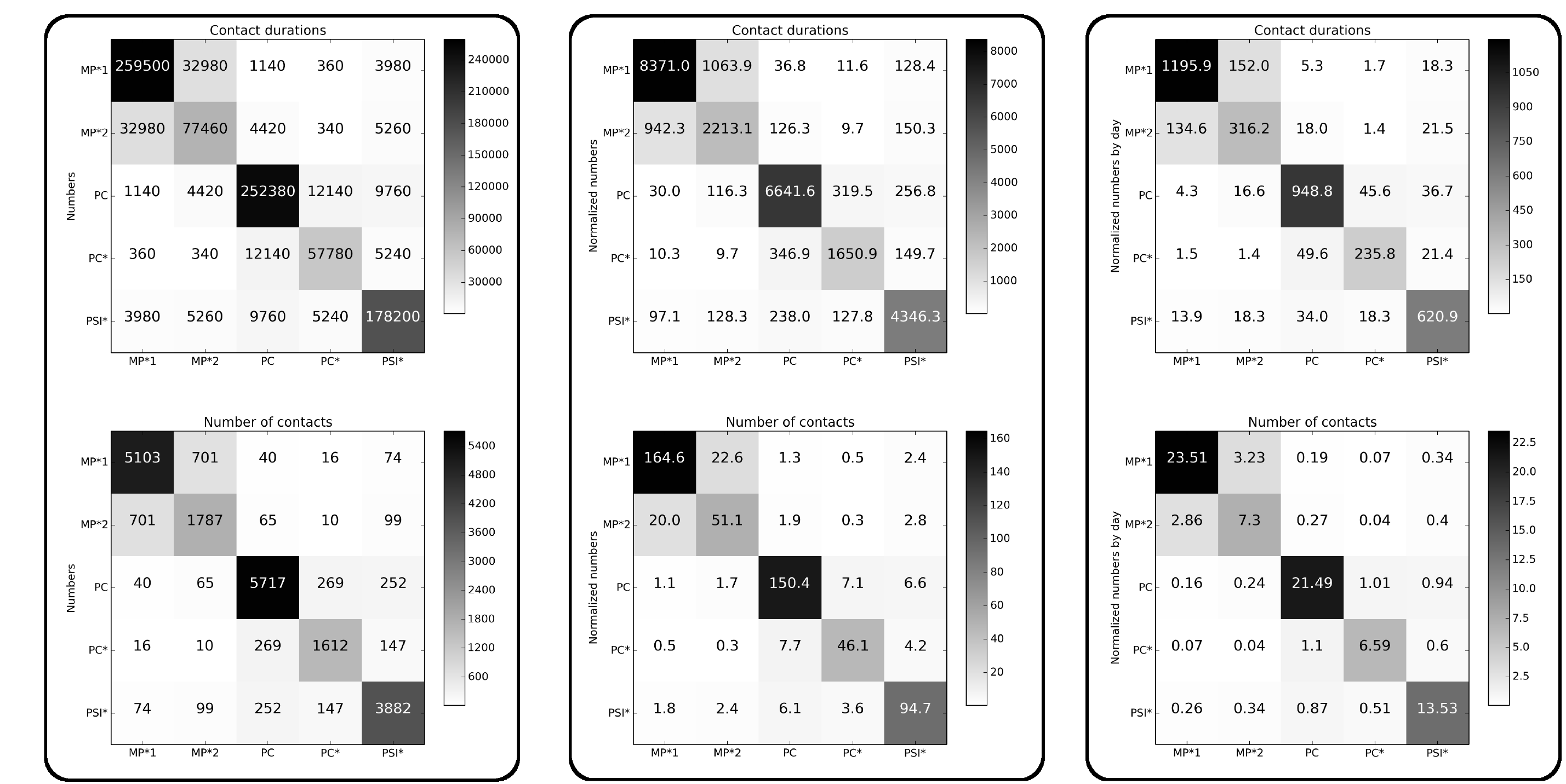}
\caption{Contacts matrices giving the cumulated durations in seconds (first row) and the numbers (second row) of contacts between classes during the whole
study.
In the first column, the matrix entry at row X and column Y gives the total duration (resp. number) of all contacts between all individuals of class X with all individuals of class Y.
In the second column, the matrix entry at row X and column Y gives the average duration (resp. number) of contacts of an individual of class
X with individuals of class Y. In the third column, we normalize each matrix element of the second column
matrices by the duration of the study, in days, to obtain
at row X and column Y the average daily duration (resp. number) of contacts of an individual of class X with individuals of class Y. }
\label{fig:CM}
\end{figure}

Figure \ref{fig:CM} reports the cumulative durations and the total numbers of contacts between classes of individuals, computed over 
the whole study duration. The second and third columns take into account the different numbers of individuals in each class, yielding
asymmetric matrices. The large values observed on the matrices' diagonals show that
most contacts  (18,101, i.e., 91.5\% of all contacts) involve students of the same class, indicating 
{ that contacts are strongly assortative with respect to class},
consistently to the results obtained in a primary school \cite{Stehle:2011b} { and as assumed in a number of complex
agent-based models of contact networks built from socio-demographic data \cite{Iozzi:2010}}. 
Very few contacts are observed between students of different classes. 
An additional substructure can however be noted: the five classes
can broadly be divided into two groups, a group of two classes (MP$^*$1 and MP$^*$2) and a 
group of three classes (PC, PC$^*$, PSI$^*$). More
contacts are observed between students of two classes in the same group than between students of two different groups.
This structure can have two origins: first, the topics studied by classes of each group are similar; moreover,
the classrooms of each group are physically close in the high school.

Notably, and despite the very different numbers of contacts,
the distributions of contact durations restricted to the contacts within a given class, or between students
of given classes, display similar statistical properties as the global distribution shown in Figure \ref{fig:pdt} (not shown): these
distributions are broad with strong fluctuations, and have the same functional shape.

\begin{figure}
\includegraphics[width=\textwidth]{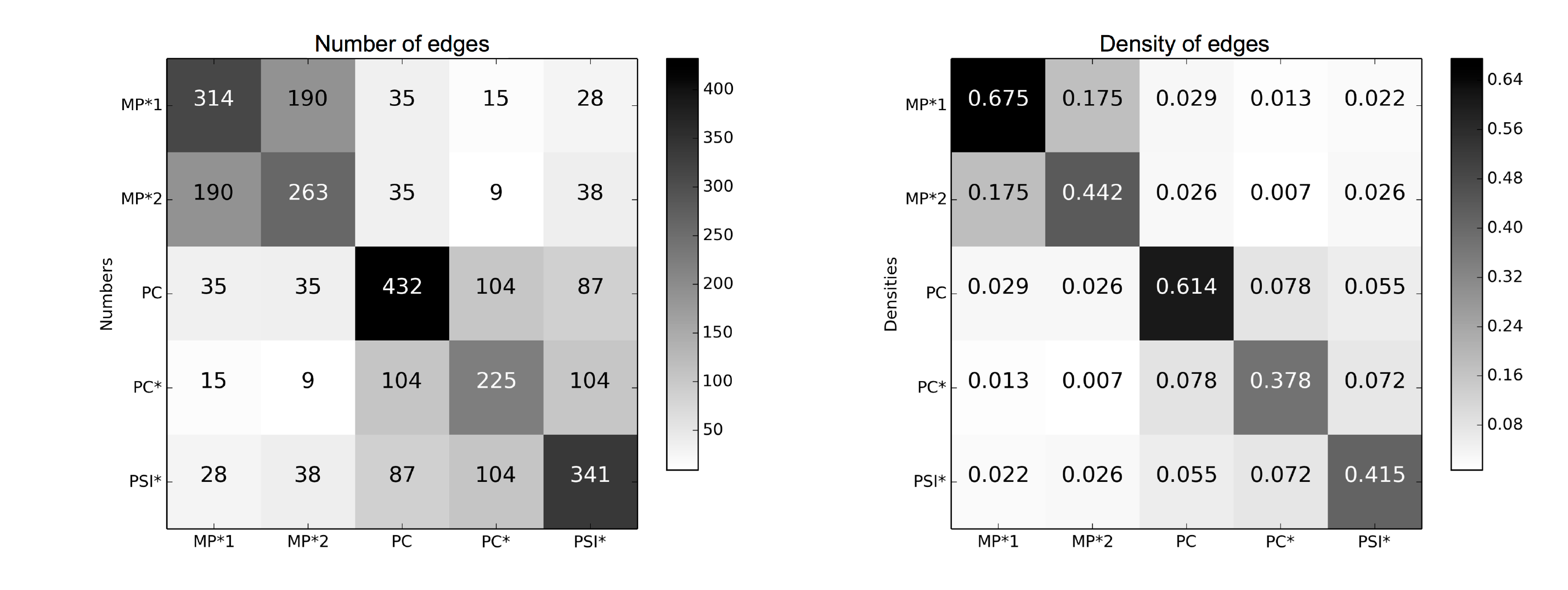}
\caption{Contact matrices of edge numbers and densities.
Left: the matrix entry at row X and column Y gives
$E_{XY}$, i.e., the number of pairs of individuals of classes $X$ and $Y$ who have been in contact at least once
during the study.
Right: the matrix entry at row X and column Y gives $\rho_{XY}$, i.e., $E_{XY}$ normalized by the maximal possible number of
pairs of individuals of classes $X$ and $Y$ ($E^{max}_{XY}=n_X n_Y$ if $X \ne Y$, $E^{max}_{XX}=n_X (n_X-1)/2$ if $X=Y$).
}
\label{fig:CM2}
\end{figure}

The contact matrices of Figure \ref{fig:CM} correspond to a coarse-grained picture in which each student is assumed to have contacts with
all other students, both within his/her class and outside, even if the average number and duration of the contacts between two individuals
vary strongly depending on their respective classes. 
Figure \ref{fig:CM2} however shows that this picture is far from correct: it gives the number of pairs of individuals of given classes who have been in contact
at least once during the study, and the same quantity normalized by its maximum possible value. For any pair of classes X and Y, this corresponds to the 
density of edges between individuals in the network of contacts aggregated over the whole study and restricted to these classes. 
This quantity would be equal to 1 if all individuals had been in contact. We observe that the density of edges is instead very small for
distinct classes, and that it is still far from 1 inside each class, even if it takes much larger values. This shows the interest of investigating
contact patterns in more detail by studying the contact network structure.

\subsection*{Contact network}

\begin{figure}[!h]
\includegraphics[width=\textwidth]{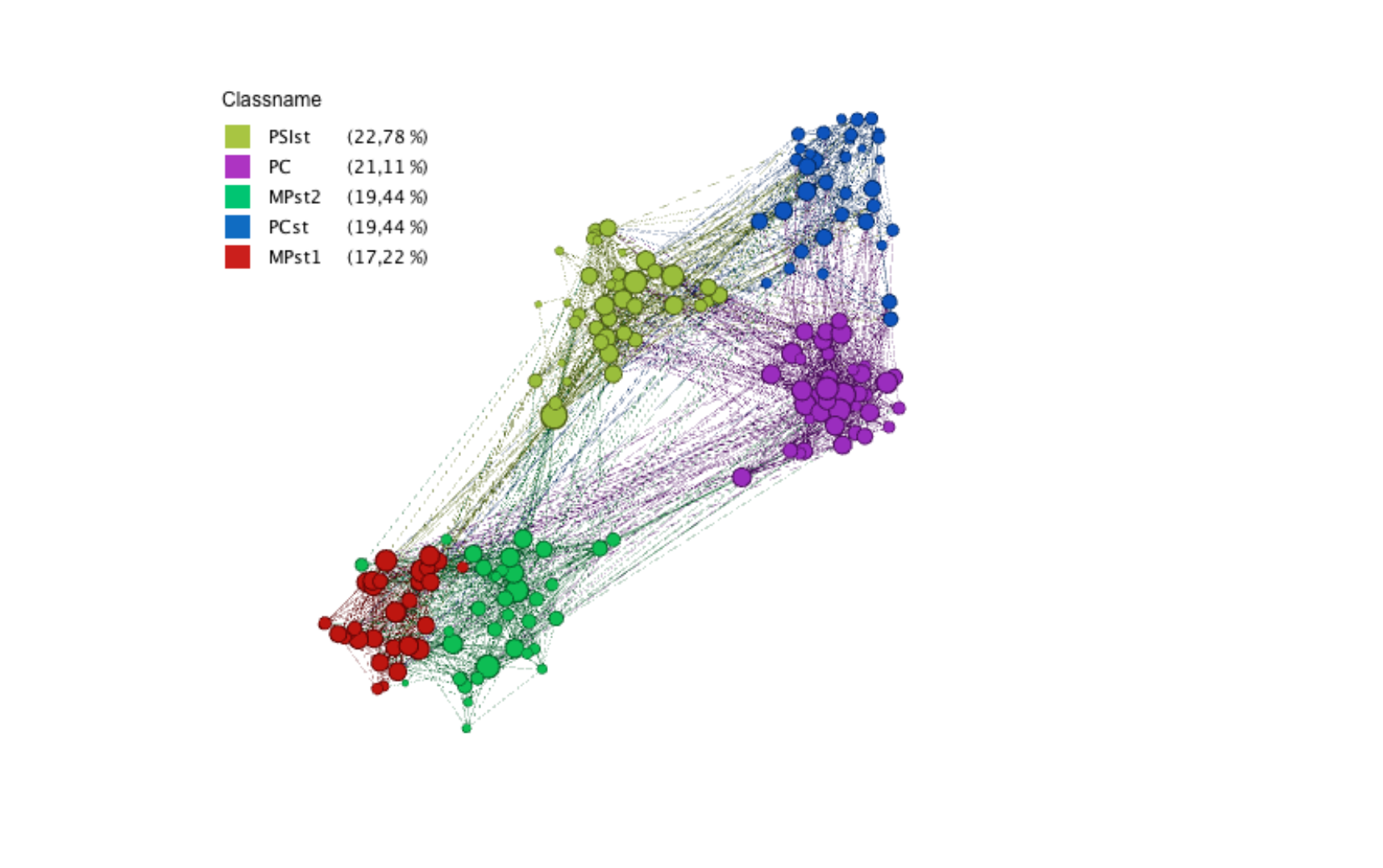}
\caption{Network of contacts between students, aggregated over the whole study duration.
Each node represents a student, its color gives the student's class and its size
is given by its degree.
Figure created using the Gephi software, http://www.gephi.org.}
\label{fig:gephi}
\end{figure}

The contact network aggregated over the whole study has 180 nodes representing the 180 students, and 2220 edges
corresponding to the pairs of students who have been in contact at least once during the data collection. The average shortest path length
in this network is equal to 2.15, and its clustering coefficient is equal to 0.48 (a random network with the same
number of nodes and edges would have a clustering coefficient $\approx 0.17$).
Moreover, the network has a strong modular structure, as already apparent from the structure of the contact matrices,
and as visible in Figure \ref{fig:gephi}. 

\begin{figure}
\includegraphics[width=\textwidth]{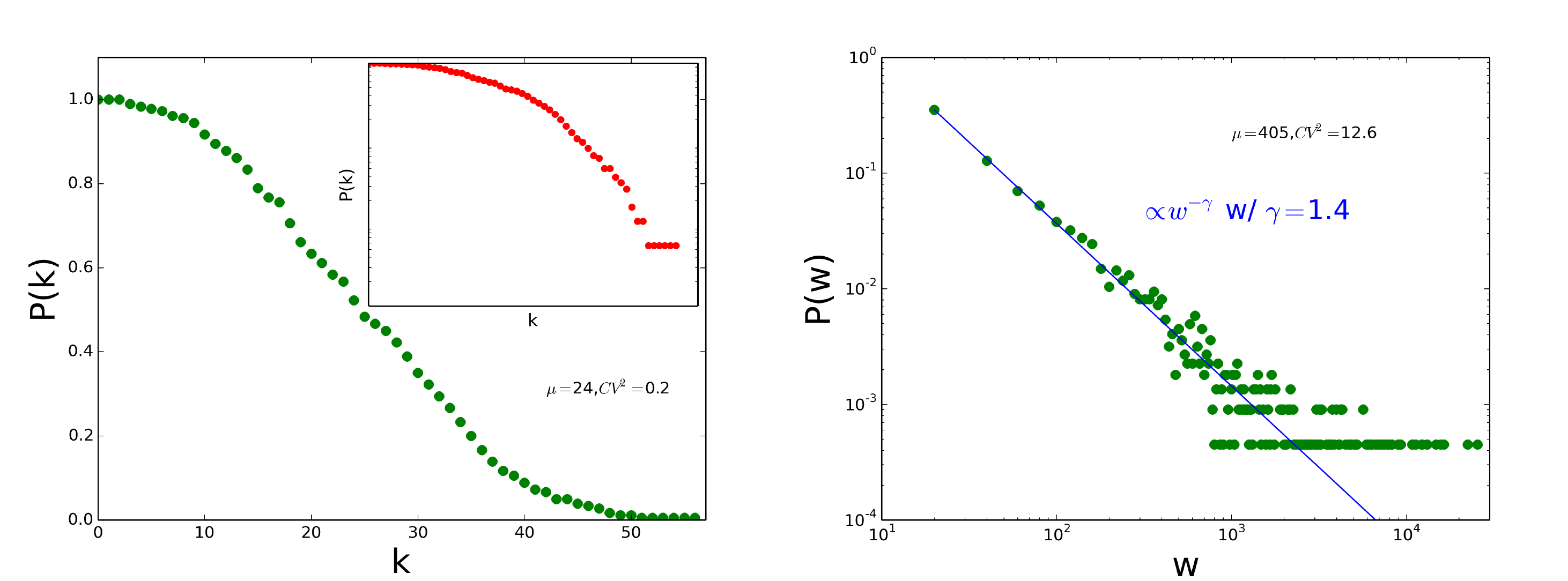}
\caption{Left: cumulative degree distribution $P(k)$ of the contact network aggregated over the whole study duration, i.e., probability that
a randomly chosen node has degree $\ge k$. Inset: the same distribution in lin-log scale.
Right: distribution of edge weights in the contact network.
The weight $w_{ij}$ of an edge i-j is given by the total time spent in face-to-face proximity by the two corresponding
individuals during the aggregation time window (here the whole study, i.e., 7 days).}
\label{fig:pkpw}
\end{figure}

Figure \ref{fig:pkpw} displays the distributions of nodes' degrees and of links' weights in the global aggregated network.
The degree distribution is narrow ($CV^2=0.2$), with a tail decaying in an approximately exponential fashion: the contact network
is not heterogeneous in terms of degrees. The narrow character of the degree distribution
of empirical human contacts has been observed as well in other contexts \cite{Salathe:2010,Isella:2011b,Stehle:2011a,Stehle:2011b}.
On the other hand, the distribution of links' weights exhibits strong fluctuations, with a squared coefficient of variation equal to 12.6
and a heavy tail that can be approximately fitted by a power law. The average  
amount of time spent in interaction by two persons is 405 seconds (6 min 45 s) during the whole study duration, but this value
hides large heterogeneities. Most cumulated durations are short (64\% of the pairs of individuals who have interacted at least once have been in contact
less than 2 minutes over the whole data collection period),
but large values are also observed: 12\% have spent more than 10 minutes in contact and 2.5\% more than 1 hour. 
Moreover, as shown in Figure S1 in File S1, the heterogeneity of cumulated contact durations is observed
both at the global population level (Figure \ref{fig:pkpw}) and if we restrict the distribution to pairs of  individuals in the same
class or in specific given classes: even within a class in which the density of edges is relatively large, as shown by the
contact matrices, different pairs of individuals can have widely different interaction durations.

\subsection*{Gender homophily}

The term homophily refers to the preference that individuals exhibit when they interact and build social ties with peers they consider to be alike.
It is a well-known feature of human behavior and has been studied in many contexts
\cite{McPherson:2001}. It is in particular known from social science studies based on surveys and direct observation
that children tend to exhibit gender preference at school, and that this gender homophily decreases during adolescence
{ (Note that gender homophily can also be described as sex assortativeness in social contacts)}.
Recently, statistical evidence of gender homophily has also been obtained in a high-resolution time-resolved data set
describing face-to-face proximity of children in a primary school \cite{Stehle:2013}. The present data set describes the interactions
of young adults in a high school context, and it is therefore of interest to investigate the possible presence of gender homophily with the
same methods.

\begin{figure}[!h]
\includegraphics[width=\textwidth]{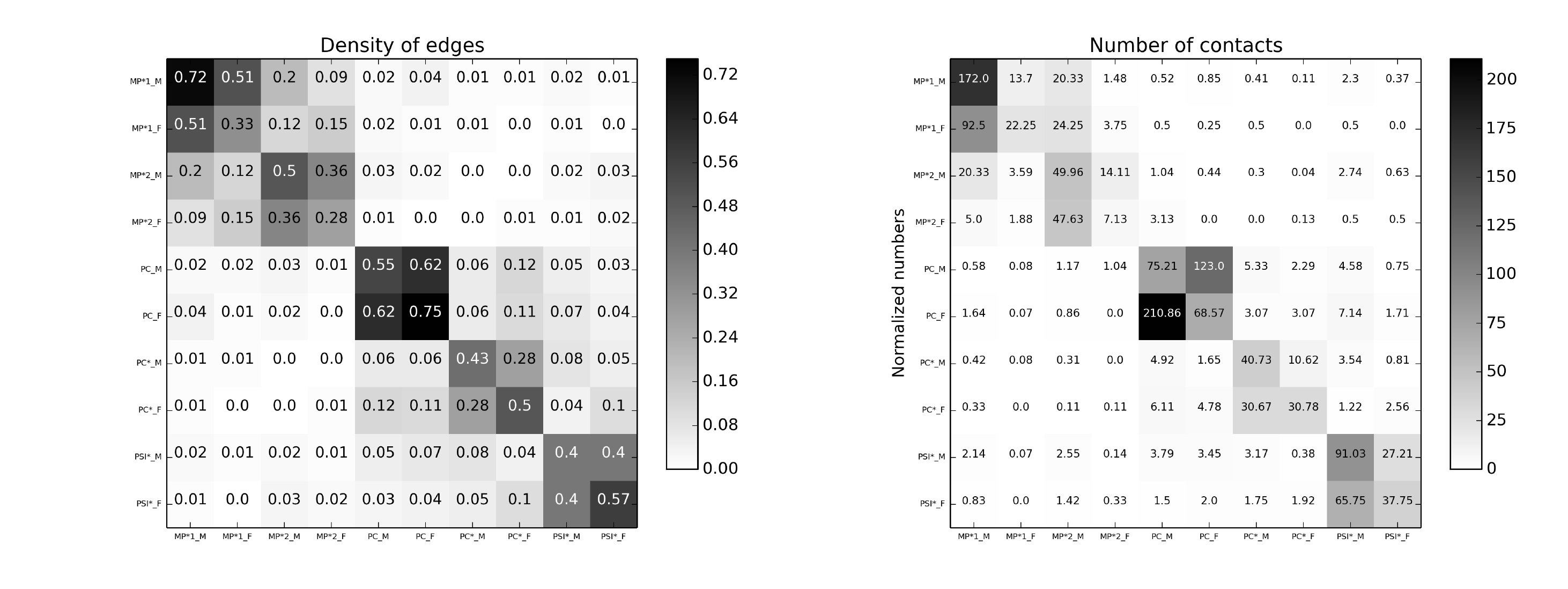}
\caption{Left: Densities of edges between groups of individuals (class + gender) of the aggregated network.
Right: Normalized numbers of contacts for the whole study duration.}
\label{fig:CMbygender}
\end{figure}

Figure \ref{fig:CMbygender} displays contact matrices giving the normalized numbers of contacts 
and the densities of edges between individuals of given groups. In these matrices, we consider 
10 groups obtained by dividing each class into two groups according to the students' gender.
As the numbers of male (M) and female (F) students are strongly different (see Table \ref{table:2}), with much more
male than female students, we consider normalized contact matrices: for the number of edges,
we normalize by the maximum number of edges between each pair of groups and, for the contact 
durations, each matrix element at row X and column Y is normalized by the number of individuals in group X
in order to give the average time spent by a member of group X with individuals of group Y.

The contact matrices display a block diagonal form, each 2x2 block corresponding to a class.
The number and durations of contacts among female students is typically low; in the 
global aggregated contact network, 57.7\% of the edges join two male students,  7.9\% join two female students,
and 34.4\% are between students of different gender.

\begin{figure}[!h]
\includegraphics[width=\textwidth]{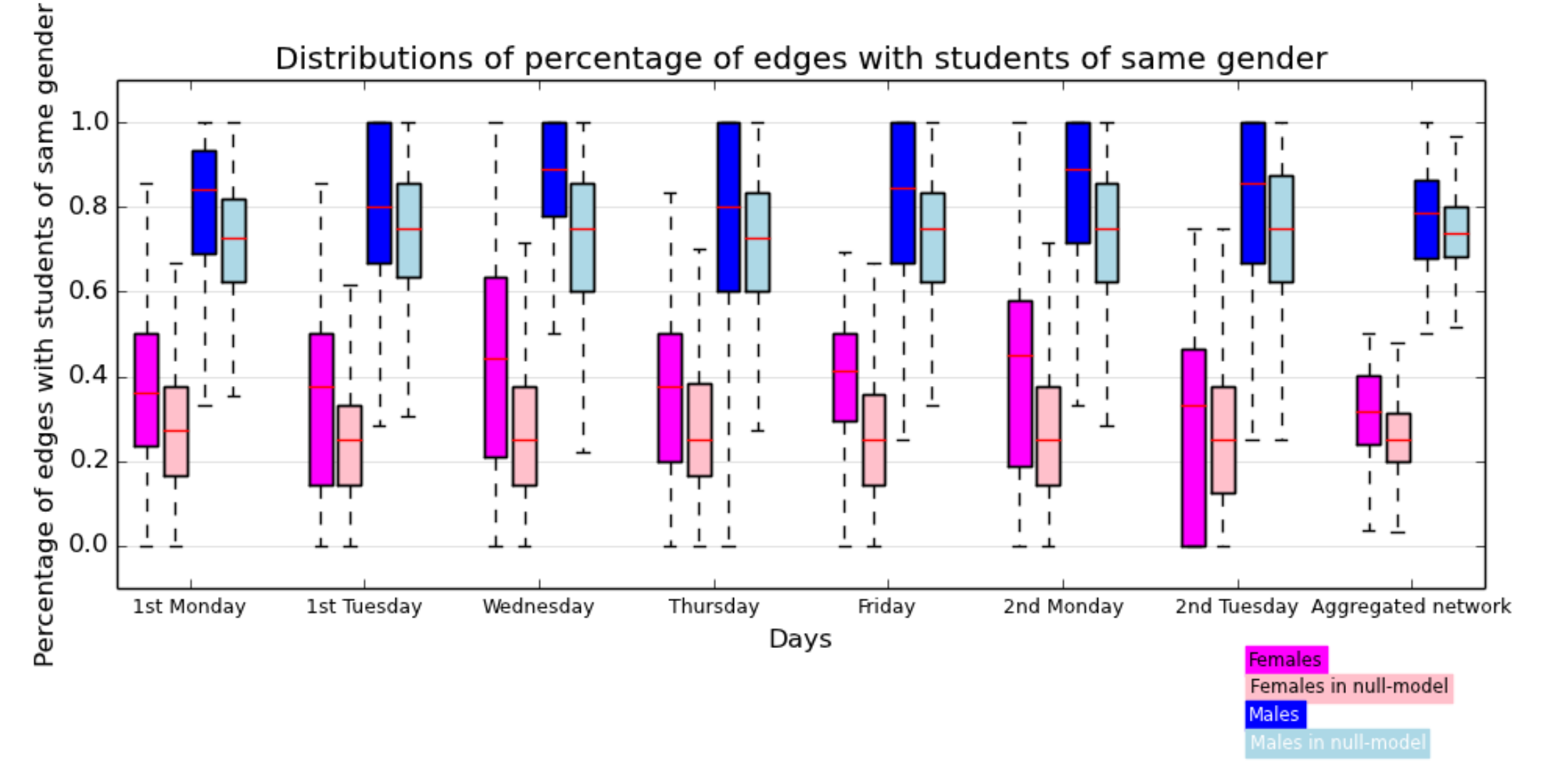}
\caption{Boxplots showing the distributions of the fraction of edges with
students of same gender for males and females with empirical data and using the null-model described above.
The centre of each box indicates the median of the distribution, its extremities the $25\%$ and $75\%$ quartiles.}
\label{fig:homophily}
\end{figure}

These values seem to indicate a preference for contacts with students of the other gender among female students and for
students of the same gender for male students. This appears
as well through the distribution of the same gender preference index $P_{sg}$: for each individual, this index is defined
as the fraction of edges, in the aggregated contact network, with individuals of the same gender. The corresponding
distributions are shown separately for male and female students as boxplots in Figure \ref{fig:homophily} for the contact 
networks aggregated on each day and
over the whole data collection. The fraction of same-gender neighbors is systematically high for male students and
lower than 0.5 for female students. The average values of these distributions are given in Table \ref{table:homophily}.

\begin{table}
\begin{tabular}{|c|c|c|c|c|c|c|c|c|}
\hline
 & 1st Mon & 1st Tues & Wed & Thurs & Fri & 2nd Mon & 2nd Tues & Aggregated\\
&&&&&&&&network\\
\hline
 Males & 80.9 & 77.8 & 84.0 & 74.7 & 78.7 & 83.7 & 80.9 & 77.6 \\
\hline
Females & 38.8 & 34.5 & 45.4 & 37.0 & 38.4 & 41.8 & 28.3 & 31.9 \\
\hline
\hline
Males, reshuffled network & 71.6 & 73.9 & 72.8 & 71.6 & 73.3 & 73.6 & 74.0 & 73.7 \\
\hline
Females, reshuffled network & 27.7 & 25.3 & 26.3 & 27.5 & 26.0 & 25.7 & 25.5 & 25.8 \\
\hline
\end{tabular}\\
\caption{Average same gender preference index for males and females for each day and in the
globally aggregated contact network, for the original data and in the null model (average over $1000$ realizations).}
\label{table:homophily}
\end{table}

To interpret these values, we need however to take into account 
the very strong imbalance between male and female students, which clearly plays an important role here. For instance, in the limit
of a very small fraction of female students, the fraction of female-female (F-F) interactions would be negligible even in the case of a 
fully homogeneous, gender-indifferent, mixing of individuals. We therefore consider a simple null model for each aggregated graph,
given by a graph with the same number of nodes and edges but randomly placed edges.
The fractions of edges joining male students is then 54.5\% on average, 
while 6.7\% of the edges join female students (averages over $1000$ realizations of the null model).
Moreover, the distributions and averages of the same gender preference
index in the reshuffled networks for each day and for the global study are shown in Figure \ref{fig:homophily} and Table \ref{table:homophily}.
The same gender preference index takes values systematically slightly {\em lower} in the null model than in the data for both male and female students. This tends to indicate a slight tendency towards gender homophily for both genders. The boxplots displayed
in Figure \ref{fig:homophily} show however that this tendency is not statistically significant, and that the observed data is in fact compatible
with a null hypothesis of absence of homophily and of gender indifference in the contact patterns of the students.
We also note that keeping only the links corresponding to a weight larger than a given threshold,
corresponding e.g., to an aggregated interaction duration of $2$ or $5$ minutes over the study duration,
changes the number of edges of each type but does not change the results concerning the absence
of gender homophily.

Another way (and reason) to assess the presence of homophily in the classes is related to the information of epidemiological models
by data on human contacts. As appears clearly from the contact matrix and contact network analysis, the population of high school 
students is far from being homogeneously mixed, and it is certainly relevant to use a level of description in which students are
divided into groups corresponding to their respective classes. If a strong gender homophily were to be observed, corresponding
to the presence of a strong group substructure inside classes, it would as well be important to consider such substructure in models
of contact patterns in order to describe spreading phenomena in such population. The assessment of such properties is thus related 
to the issue of the amount of information needed to inform models, as discussed in \cite{Blower:2011,Stehle:2011a,Machens:2013}.
While a detailed analysis of spreading processes in the population under study goes beyond the scope of the present investigation,
we investigate this point in the Supplementary File S1 through numerical simulations of a simple Susceptible-Infected
spreading process: we show that 
a description of the contact patterns at the level of classes corresponds to a sufficient resolution, and that
it is not necessary to represent the students population at the finer level of a division into 
gender groups in each class.

\subsection*{Longitudinal analysis}


We now turn to the longitudinal study of the contact patterns. Of particular interest are the potential similarities between the properties of contacts
occurring in different days, in order to understand for instance how much information is lost if data are gathered only during one single
day or few days, and how much data gathering is needed to inform models of human behavior. 

\begin{figure}[!h]
\includegraphics[width=\textwidth]{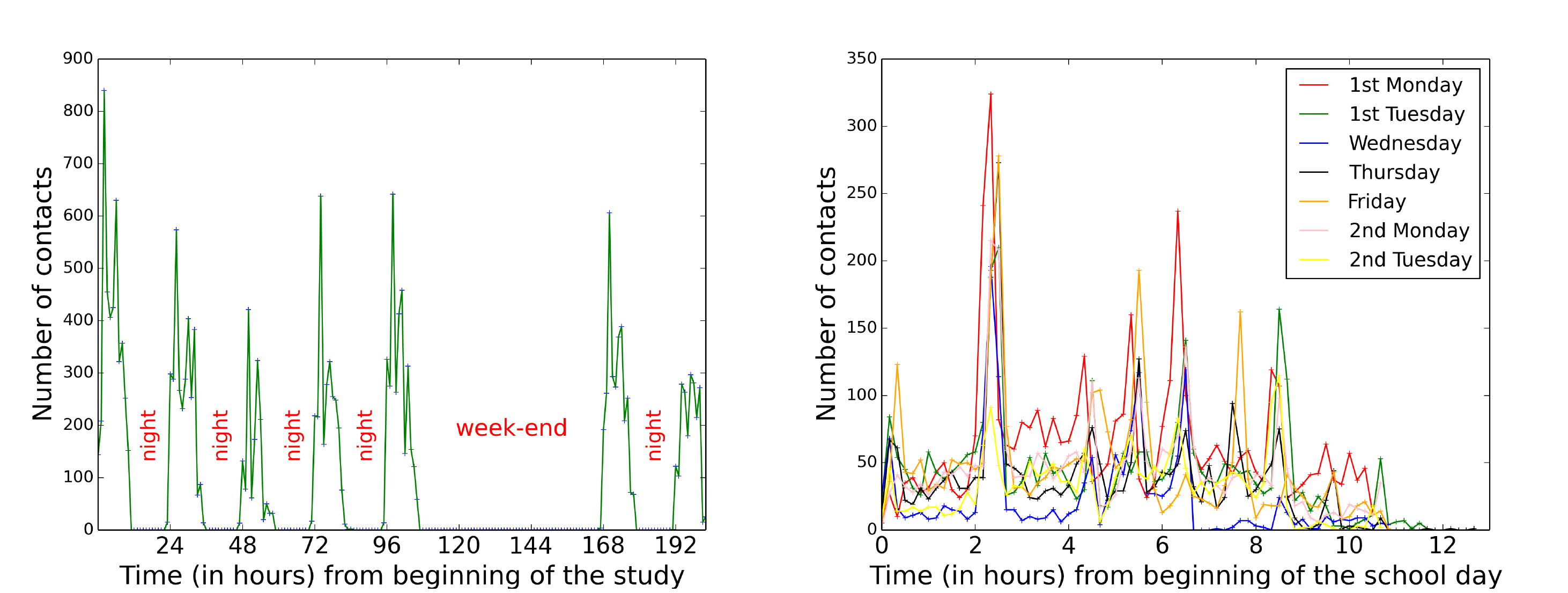}
\caption{Left: number of contacts per one-hour time-windows. Right: number of contacts per 10-minutes periods for each day.}
\label{fig:timeline}
\end{figure}

Figure \ref{fig:timeline} reports the evolution of the number of contacts at the temporal resolutions of one-hour and ten-minutes time windows
{ (see also Supplementary File S1)}. 
The number of contacts fluctuates strongly over the course of each day, { with strong peaks determined by the various breaks between lectures,}
showing that a data collection limited to a few hours
would be insufficient to gather an accurate picture of the contacts occurring in the high school; however, the evolution is very 
similar from one day to another, with maxima determined by the lunch and class breaks. { We also note that Wednesday
afternoon has much less contacts than other days, due to the fact that students pass exams which typically last the whole afternoon without any
break.}

Contact matrices defined for each day, each morning (before 1PM) and each afternoon (after 1PM) 
periods are displayed in Figures S3 and S4 of File S1. While the precise values of the numbers (and cumulative durations) 
of contacts between classes fluctuate from one day to the other and from a half-day to another, the 
structure of the contact matrices presents a robust pattern, with higher values on the diagonal and a
substructure of two groups of 2 and 3 classes, as observed for the contact matrix aggregated over the whole study duration.
In order to quantify this point more precisely, we compute the cosine similarities between (i) pairs
of daily contact matrices and (ii) morning and afternoon contact matrices for each day. The resulting values,
given in Table S1 of File S1, are large, with a minimum of 0.65 and a maximum of 0.99 between daily
contact matrices (with most values above 0.9), and values between 0.53 and 0.94 between morning and afternoon contact matrices. 
In order to check if these high values are only due to the fact that the diagonal elements of the contact matrices take much larger values than the 
off-diagonal ones, we also compute similarities restricted to the off-diagonal elements of the matrices. The corresponding values are still very
large, with a minimum at 0.56 and a maximum of 0.97, and most values above 0.8.
This shows that, despite the fluctuations in the number of contacts, the structure of the contacts between classes is very robust 
across different days, and is well captured by a data collection performed on any given day.

\begin{figure}[!h]
\includegraphics[width=.8\textwidth]{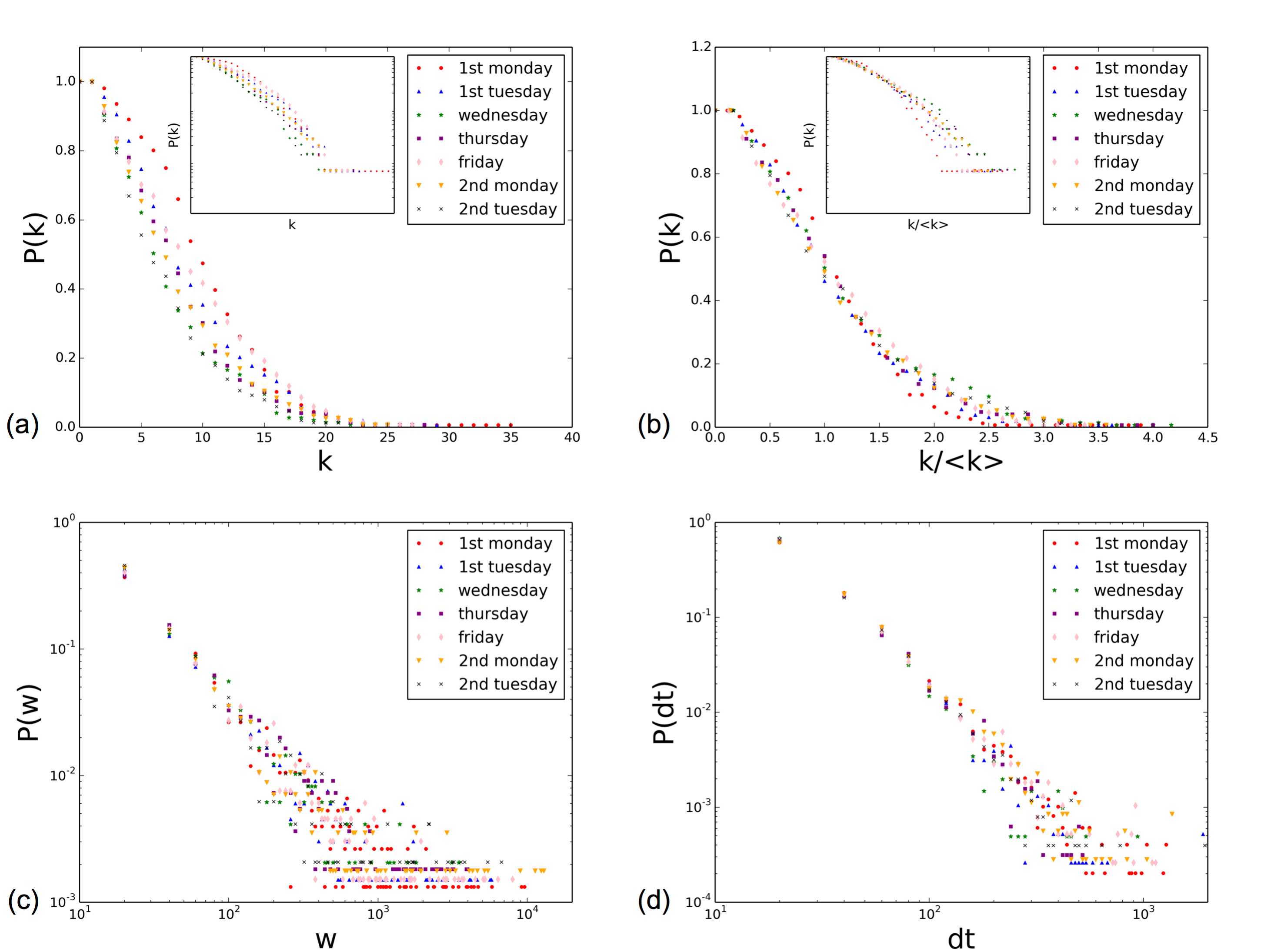}
\caption{Properties of daily aggregated network. (a) Cumulative degree distribution of the daily aggregated networks.
(b) Cumulative distribution of degrees rescaled by the average degree $<k>$ of each daily network. The insets show the
same data in lin-log scale. (c) Distribution of edge weights for each daily network. (d) Distribution of contact durations in each day.}
\label{fig:pkpwpdt_day}
\end{figure}

In order to shed more light into this robustness, we also
compare the contact networks aggregated over daily time windows. 
From the point of view of the statistical properties, Figure \ref{fig:pkpwpdt_day} highlights  the similarity between the
statistical distributions of node degree and contact durations, both for the durations of single contact events and for the daily cumulative
durations of contacts between individuals. In the case of the degree distributions (shown separately in Figure S5 in File S1), 
slightly different average values are observed in different days,
but the functional shapes of the distributions are similar, and the rescaled distributions are very close. The distributions of contact durations and
of link weights are as well on top of each other, and fitting by a power law yields similar exponents (see Figures S6 and S7 in File S1).

\begin{figure}[!h]
\includegraphics[width=.8\textwidth]{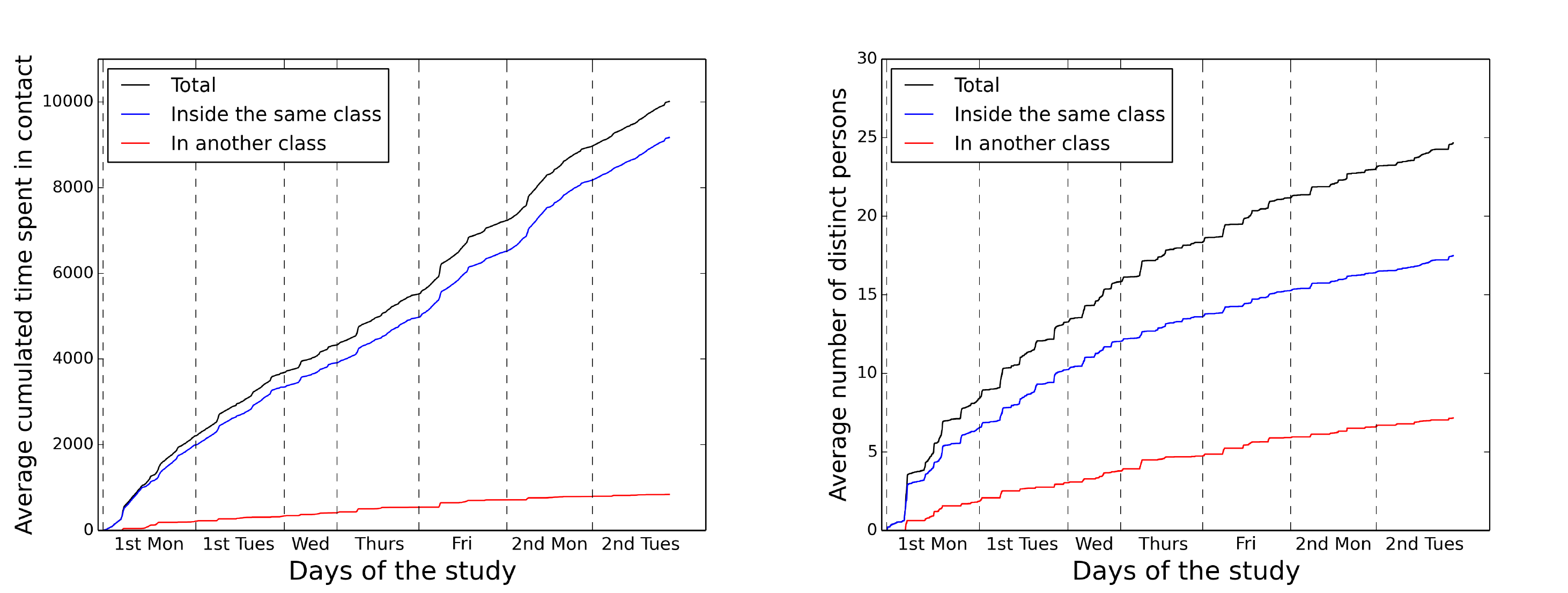}
\caption{Left: time evolution of the average cumulated time spent by a student in contact with other students during the study. The average total time is displayed in black, the average time spent with students of the same class in blue, and the average time spent with students of other classes in red.
Right: time evolution of the average number of distinct persons with whom a student has been in contact since the start of the data collection.
The average total number is displayed in black, the average number of individuals of the same class in blue, and the average number of persons of other 
classes in red. }
\label{fig:deg_vs_time}
\end{figure}

The strong similarity in the statistical properties of the daily contact networks is however not the whole story, as
Figure \ref{fig:deg_vs_time} shows. On the one hand, 
the total time spent in contact by an average student grows regularly over time, both with students of the same class and with students of
different classes, showing that the amount of time spent in contact each day by a student does not fluctuate strongly from one day to the next,
as also observed in a primary school \cite{Stehle:2011b}.
On the other hand, the average number of distinct individuals with whom a student has been in contact displays 
a strictly increasing behavior over the whole study duration, with no clear saturation trend { (Note that  Fig. \ref{fig:deg_vs_time} 
shows averages: at each time, the number of distinct persons contacted by a student has a distribution similar to the one shown in Fig. \ref{fig:pkpw}),
and this distribution shifts towards larger values while retaining its shape as time increases.} 
This means that an average student 
continues to meet new persons each day, and that his/her neighborhoods in the contact network, { i.e., his/her individual
contact patterns},
 change from one day to the next.
The figure also shows that students meet a larger number of distinct individuals of the same class than of other classes, as expected from
the previous analysis of contact matrices and networks, and that both numbers continue to grow during the whole study. 

\begin{figure}[!h]
\includegraphics[width=.8\textwidth]{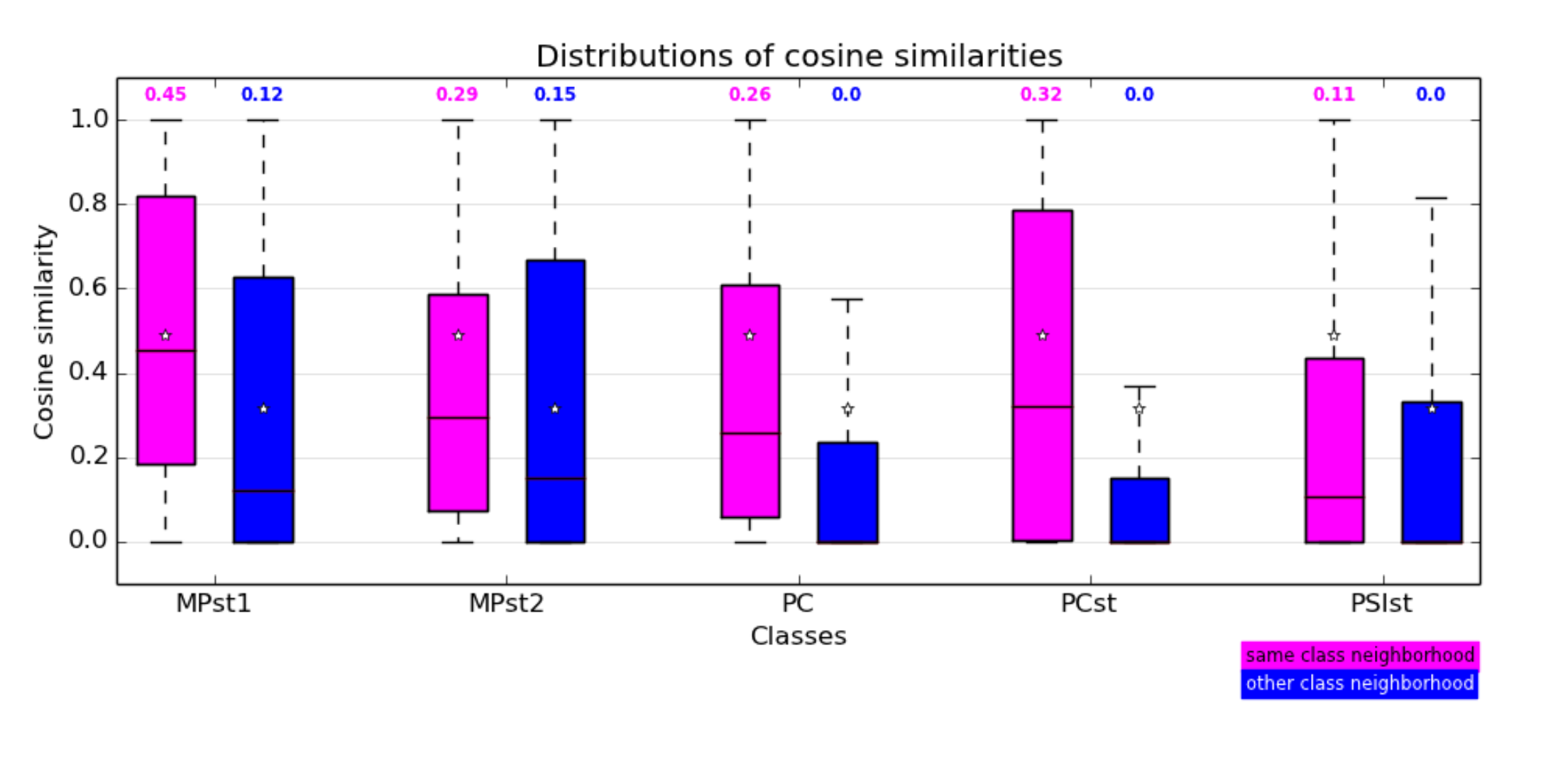}
\caption{Boxplots showing the distributions of cosine similarities of neighborhoods of nodes in pairs of
daily contact networks, for each class, and restricting the neighborhoods to intra-class (violet) and inter-class (blue) neighborhoods.
These distributions are obtained in the following way: in each class, for each person we calculate the cosine similarities of his/her neighborhood inside
the class and outside the class for each pair of days.
Each distribution thus corresponds to 21(couples of days)*N(number of students in the class) similarity values. The center of each box gives the median
(value given above each box) and its extremities correspond to the 25\% and 75\% quartiles.
The star symbols show the mean value of each distribution.}
\label{fig:net_similarities}
\end{figure}

In order to quantify the change in the students' neighborhood in different days, we compute 
the cosine similarities between the neighborhoods of each node in each pair of daily aggregated contact networks. 
The distributions of these similarities are shown in Figure \ref{fig:net_similarities} for students of each class, distinguishing
between the neighborhoods of each student inside his/her class and with students of another class. Cosine similarities restricted to
intra-class neighborhoods tend to be larger than the ones restricted to interclass neighborhoods, indicating a slightly
larger stability of intra-class neighborhoods. The averages are given in Table S3 of File S1. 

The values of the cosine similarities observed are rather far from $1$, indicating substantial changes of individual contact patterns
across days.
In order { however} to better grasp the meaning of these values, {i.e., to understand if they ought to be considered 
"small" or "large"}, we compare the empirical values to the ones obtained with several null
models. We first consider null models in which the network edges are placed at random between the nodes of the networks, either
(a) with no restriction or (b) such that edges joining nodes of classes X and Y are randomly assigned to pairs of nodes of classes X and Y, i.e.,
maintaining the overall class structure of the network. We also consider edge rewirings which conserve the degree of each node
(``Sneppen-Maslov'' null model \cite{Maslov:2004}),
once again either (a) globally or (b) separately for each class pair. Finally, we keep the network topology unchanged
but we reshuffle the weights on the edges, either (a) reshuffling the weights of all edges or (b) reshuffling separately the weights of edges between 
each pair of classes. 

\begin{figure}[!h]
\includegraphics[width=.8\textwidth]{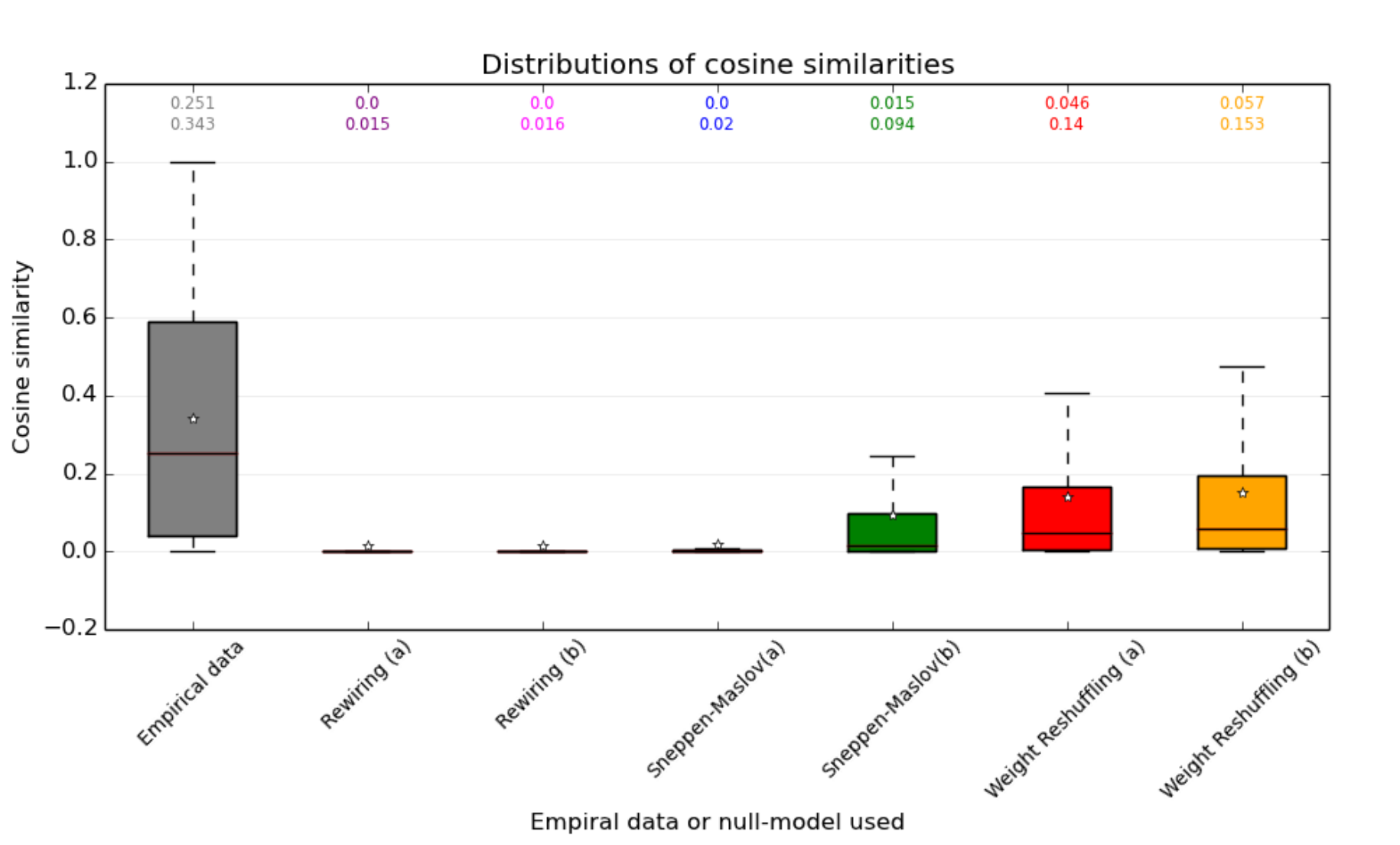}
\caption{Boxplots showing the distributions of similarities of nodes' neighborhoods in different days (for all pairs of days in the data set),
for the empirical data and for the various null models ($1000$ realizations for each).
The centre of the box gives the median, its extremities the 25\% and 75\% quartiles,
and the whiskers are at distance 1.5 times the interquartile range from the extremities (with saturation at 0 and 1). The
first number above each boxplot gives the median of the distribution and the second number gives the average, also shown by a star.}
\label{fig:sim_nullmodels}
\end{figure}

The averages of the cosine similarities obtained in each null model are given in Tables S5, S6 and S7 of File S1, 
and the distributions are shown
as boxplots in Figure \ref{fig:sim_nullmodels}. The empirical values of the cosine similarities are larger 
than the ones obtained with the null-models, even with the last null-model in which the topological structure of the contact network
and the statistical properties of the cumulative contact durations
are kept at the level of each class. This comparison with a set of null models allows us to determine that the similarities
of students' contact patterns from one day to another, { although they can seem not to be very large in absolute terms, 
are much higher than what could be expected under null hypothesis: in other terms,}
changes  in the contact neighborhoods of students from one day to another are substantial, but much less than in a situation in which
contacts would occur at random, and even less than if 
the cumulative contact durations were attributed at random on a fixed topological structure of the contact networks.
This emphasizes the need to take this robustness of contact patterns into account in models of contacts between individuals, as 
done e.g., in \cite{Stehle:2011a}.

\subsection*{Long term stability of contact patterns: comparison between data collected in different years}

We take here advantage of the fact that data were collected in the same context in two different years
to investigate the long term stability of the contact patterns between students in the high school. 
As students participating in the data collection in both years were not the same, we cannot study the change in 
individual behaviors, but focus on the overall structure of the contact networks and matrices. 


Table \ref{net_years} compares the main statistics of the 
aggregated contact networks of 2011 and 2012, both at the global level and for each class. Despite fluctuations in the absolute values,
which can be expected as the data concern different sets of individuals and different durations, similar properties are observed,
with high values of the clustering coefficient and short average paths lengths. Higher edge densities are also observed within each class,
consistently with the strong class structure observed previously. 

\begin{table}
\begin{tabular}{|p{3cm}|p{2cm}|p{2cm}|p{2cm}|p{2cm}|p{2cm}|p{2cm}|}
\hline
 &\multicolumn{3}{c|}{\Large\textbf{2011}}&\multicolumn{3}{c|}{\Large\textbf{2012}}\\
\hline
Number of nodes  &\multicolumn{3}{c|}{126}&\multicolumn{3}{c|}{180} \\
\hline
Number of Edges &\multicolumn{3}{c|}{1,710}&\multicolumn{3}{c|}{2,220}\\
\hline
Number of contacts &\multicolumn{3}{c|}{10,432}&\multicolumn{3}{c|}{19,774}\\
\hline
Cumulative duration of all contacts &\multicolumn{3}{c|}{561,010 (156 hours)}&\multicolumn{3}{c|}{900,940 seconds (250 hours)}\\
\hline
\hline
&Clustering coefficient & Average shortest path length&Density&Clustering coefficient &Average shortest path length &Density\\
\hline
Total &0.58&1.95&0.22&0.48&2.15&0.14\\
\hline
PC&0.72&1.39&0.62&0.73&1.39&0.62\\
\hline
PC$^*$&0.75&1.39&0.61&0.53&1.66&0.38\\
\hline
PSI$^*$&0.66&1.51&0.49&0.61&1.62&0.41\\
\hline
MP$^*$1&&&&0.77&1.32&0.68\\
\hline
MP$^*$2&&&&0.60&1.61&0.44\\
\hline
\end{tabular}\\
\caption{Comparison of the properties of the global aggregated networks of the 2011 and 2012 data collections.
}
\label{net_years}
\end{table}

\begin{figure}[!h]
\includegraphics[width=.8\textwidth]{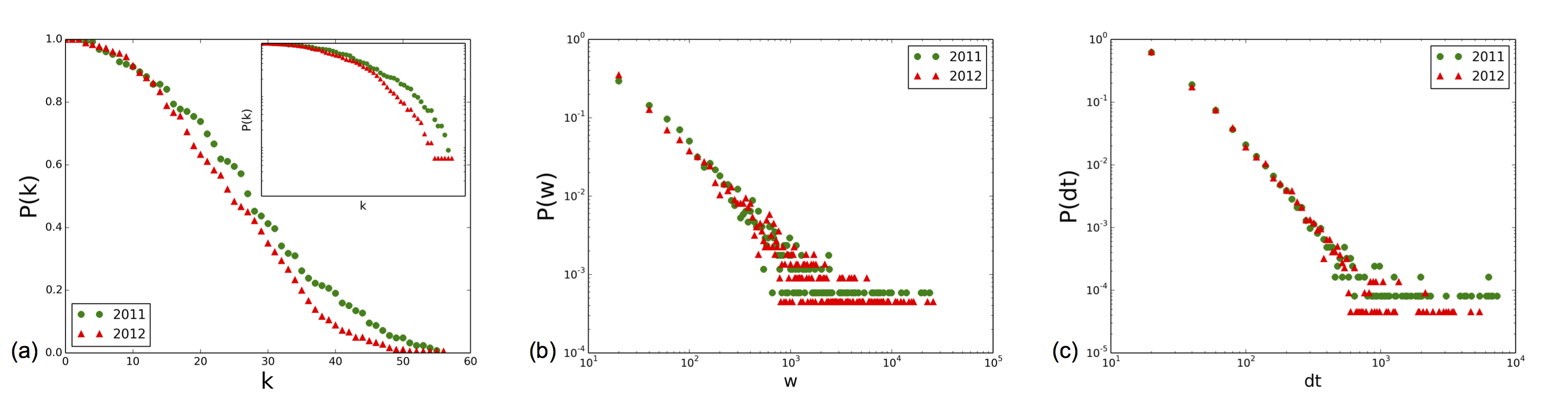}
\caption{Distribution of degrees (a), distribution of weights (b) and distribution of contact durations (c) for 2011 and 2012.}
\label{fig:pk_years}
\end{figure}

Figure \ref{fig:pk_years} 
displays the distributions of node degrees and link weights of the contact networks obtained in 2011 and 2012,
aggregated over the whole data collection duration, as well as the distributions of contact durations. All distributions
are very similar, with an exponential decrease at large degree values for the degree distributions, and very broad 
weights and contact duration distributions which collapse on top of each other for the data obtained in the different years.

\begin{figure}[!h]
\includegraphics[width=\textwidth]{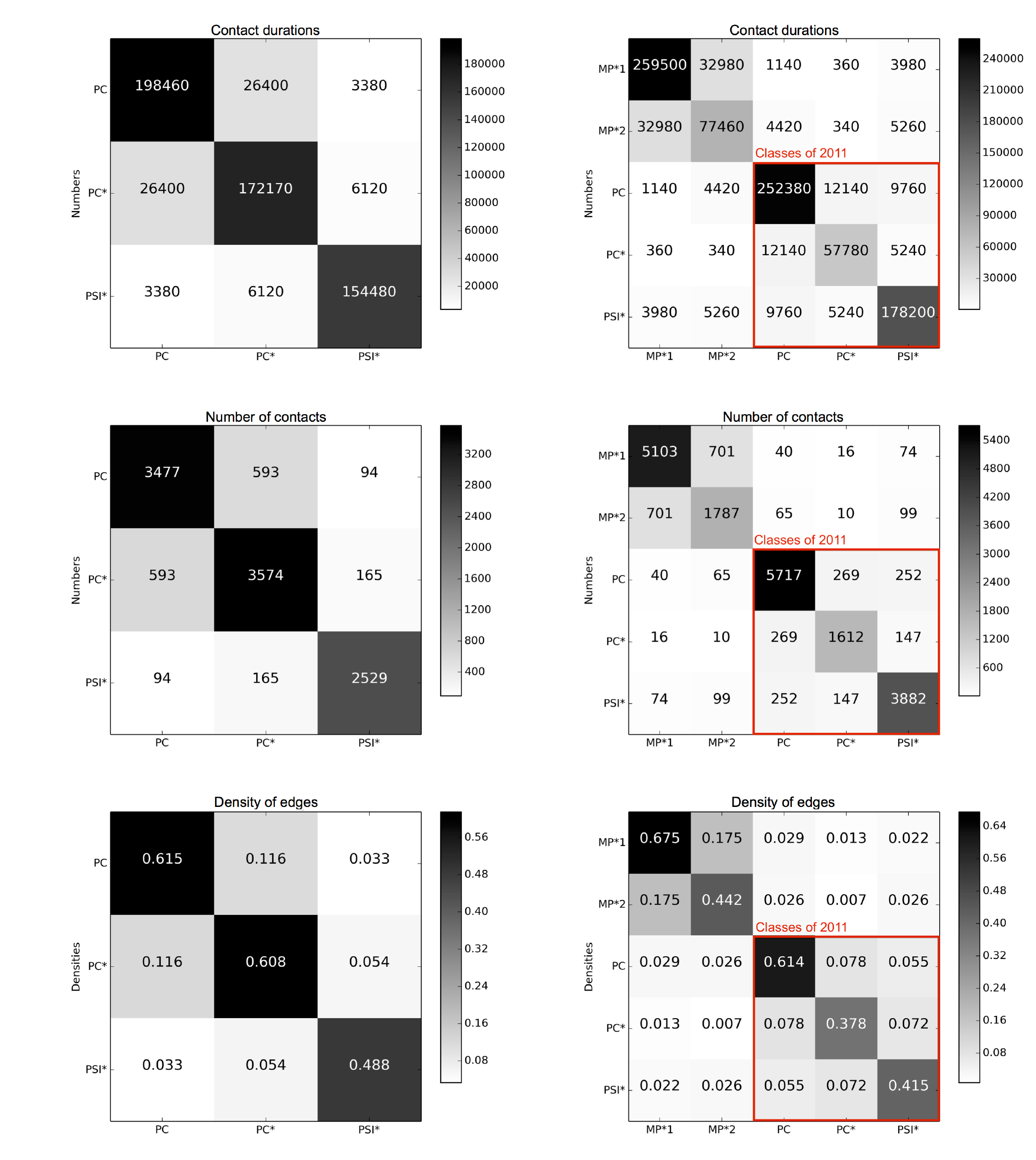}
\caption{Contact matrices giving the cumulated durations of contacts (1st line), the number of contacts
(2nd line) and densities of edges between classes (3rd line) for 2011 (1st column) and 2012 (2nd column).}
\label{CM_years}
\end{figure}

Figure \ref{CM_years} moreover displays the contact matrices describing the structure and numbers of contacts between classes.
As the data collection of 2012 involved more classes than the one of 2011, 
we evaluate the similarity between the contact patterns of two years by computing the similarity between the matrices 
collected in 2011 and the contact matrices restricted to the three classes PC, PC$^*$ and PSI$^*$ of 2012. We obtain very high values:
91.2\% for the matrices of contact durations, 92.4\% for the matrices giving the number of contacts, and 97.9\% for 
the matrices of the densities of edges between classes.


\begin{figure}[!h]
\includegraphics[width=\textwidth]{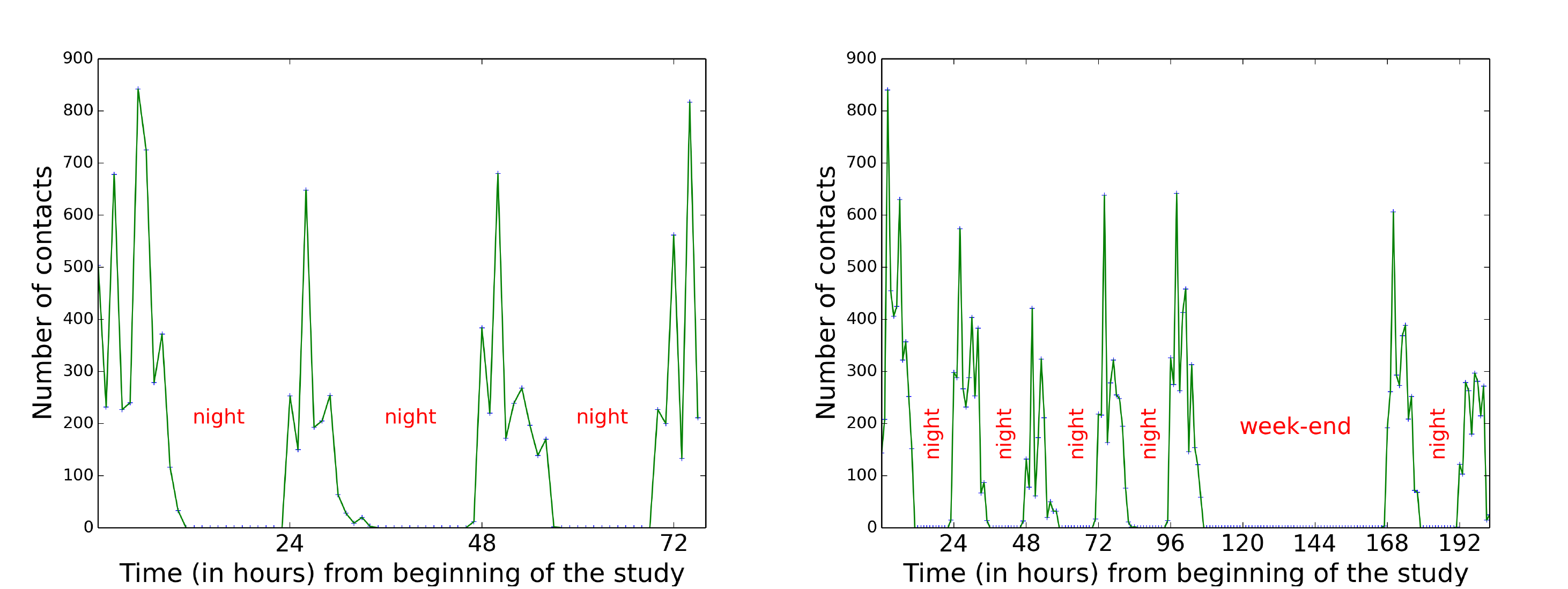}
\caption{Number of contacts per one-hour periods for the 2011 (left) and 2012 (right) data sets.}
\label{fig:timeline_years}
\end{figure}

The overall contact structures are therefore very robust from one year to the next, despite the different populations involved.
In order to investigate in more detail these similarities between years, we show in Figure \ref{fig:timeline_years}
the temporal evolutions of the number of contacts registered in one-hour time windows in both cases. The number of contacts
vary strongly within each day in both years, but the temporal patterns are very similar  from one day to another
in both cases, with daily rhythms due to class and lunch breaks. 
In both years however, the contacts of each individual in different days are not completely the same, as shown above in Figure \ref{fig:deg_vs_time}
and through the measure of the cosine similarities of neighborhoods in different daily aggregated networks. 
Interestingly, the distributions and average values of these cosine similarities take similar values in both years:
for instance, the average values of the cosine similarities of individual neighborhoods in different days vary from 0.35 to 0.43 in the 2011
data set  and from 0.29 to 0.44 in the 2012 data set. The rates of renewal of the contact neighborhood of each individual is thus as well
a robust property of these data sets.


\section*{Discussion}

In this article, we have presented an analysis of two high-resolution contact
data sets collected in a French high school using wearable sensors, respectively for three classes
during four school days in December 2011 and for five classes during seven school days in November 2012.
 
\begin{figure}[!h]
\includegraphics[width=\textwidth]{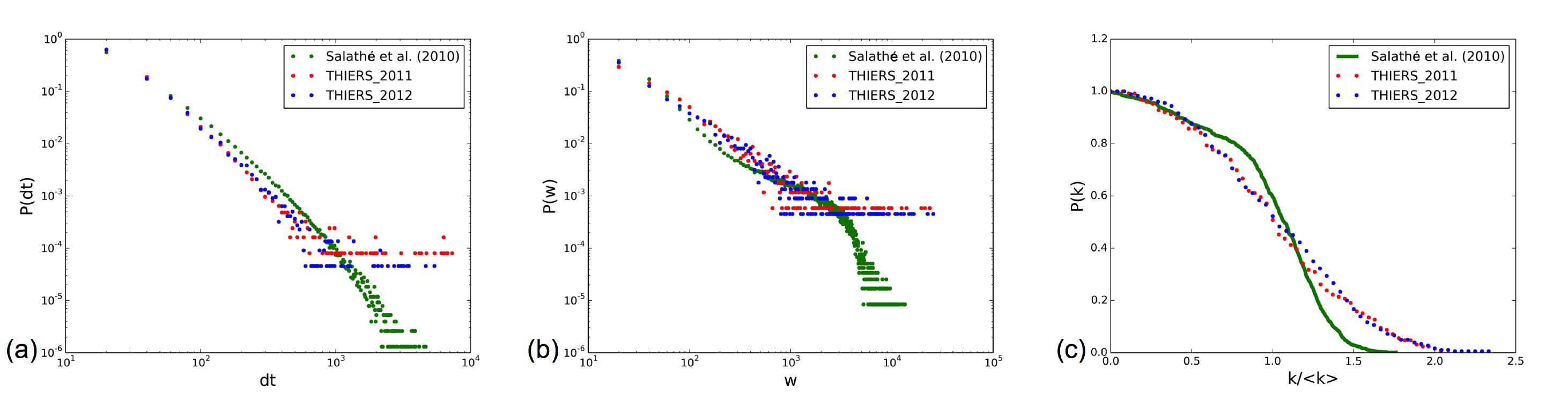}
\caption{Comparison of the distributions of (a) durations of contact events, (b) cumulated contact durations, (c) degree in the
aggregated network, for the data sets analyzed here and the data set of Ref. \cite{Salathe:2010}.}
\label{fig:salathe}
\end{figure}
 
 Several important points emerge from our analysis. The aggregated network of contacts has small
 diameter and high clustering, with a strong community structure determined by the students' classes, similarly to 
 what was observed in a primary school \cite{Stehle:2011b}. The degree distribution is homogeneous, meaning that
 the number of distinct persons with whom an individual has contacts does not show large fluctuations. On the contrary, and
 as observed in other contexts ranging from primary school to conferences or hospitals \cite{Stehle:2011b,Isella:2011a,Isella:2011b,Vanhems:2013},
 large variations occur in the durations of single contact events as well as in the cumulated durations of the contacts between two individuals:
the distribution of these quantities are heavy-tailed and no specific duration timescale can be identified. 

The contact matrices quantifying the number or duration of contacts between different classes show a clear structure,
with much larger values on the diagonal (corresponding to contacts within each class) and an additional substructure
of two groups of respectively two and three classes. While the large values of the number of contacts inside each class
is expected due to the school structure and schedule, the off-diagonal structure reflects patterns which are more due to either spatial
arrangements of classes inside the high school or to similarities in the dominant subjects studied by the students.
Overall, and as could be expected, the contact patterns in such an environment
would not be correctly represented by a homogeneous mixing approximation. 
As no strong gender homophily is observed (contrarily to the case of a primary school \cite{Stehle:2013}), a further subdivision
of each class in groups according to gender does not seem however a necessary step in modeling processes: 
the division of the population at the level of classes appears as an adequate
level of description, for instance when designing a model of contacts to evaluate
the outcome of a spreading process in this population \cite{Iozzi:2010,Gemmetto}.
 
In the context of the design of data-driven realistic models for human contacts, or for the information of models of spreading processes, 
the robustness of contact patterns at different timescales represents also a crucial information. 
Strikingly, the properties of the temporally resolved contact patterns are extremely robust from one day to the next
in several aspects: in terms of the variation of the contact numbers with time along the course of each day, 
in terms of statistical distributions of contact durations,
and in terms of the contact matrices, which have a very high similarity from one day to another. 
Interestingly, this similarity between contact patterns in the high school remains very strong when comparing data collected
in two different years, with different students populations.

The contact patterns of each individual are however not the same in different days, and we give an estimation
of the renewal of contacts by measuring the average cosine similarity of their neighborhoods in daily contact networks. 
The obtained values show that the contacts of a random individual vary substantially in different days, but much less 
 than in null models in which contacts are renewed at random from one day to the next. Moreover, the observed
 values of these similarities are similar in the two data sets corresponding to two different years, and can thus 
 be considered as a feature to be included in realistic models of human contacts in such an environment.
 
Comparison of data collected in different environments of similar nature is also important, in particular to highlight similar 
 patterns in specific structures and therefore to inform mathematical models.
Few studies using wearable sensors, and giving access to high-resolution data on contacts between individuals, are however
available. We compare in Figure \ref{fig:salathe} some statistical properties of the data sets presented here with the
data set made public by Salath\'e et al. in Ref. \cite{Salathe:2010}, which gives the durations of close proximity events between 
788 high school students during a typical day at school. Given the different definitions of contact in the two studies, the average
number of distinct neighbors in the aggregated contact networks are different:
24.7 for the present study and 299.5 for \cite{Salathe:2010}. Once rescaled however, the distributions
of degrees of the aggregated networks are similarly short-tailed, albeit with slightly different functional shapes. Most importantly,
the distributions of the contact durations and of the edge weights in the contact networks are very similar in the two studies, with
similar slopes and heavy-tails.

Data sets such as the ones presented here are of interest in various fields, including in social sciences in order to better understand and model
human behavior and interactions in different contexts, and in epidemiology in order to inform models describing the spread of infectious diseases.
In this respect, data gathering in 
contexts such as schools, where many contacts are expected between students, favoring the possible spread of diseases among them
and successively in the community, are particularly invaluable. Such data may also help 
devise and evaluate data-driven containment strategies.
For instance, the fact that the contact matrices 
are highly structured, with most  contacts occurring within classes and additional structures of 2 or three classes, as also
in a primary school \cite{Stehle:2011b}, indicates that containment strategies such as the closure of single classes or groups
of classes when a disease outbreak is detected could represent viable alternatives to full school closures
\cite{Gemmetto}.

The use of wearable sensors appears as an appealing method to measure contacts between individuals.
{ It has both advantages 
and limitations with respect to more traditional approaches such as the use of survey or time-use data 
\cite{Mossong:2008,Mikolajczyk:2008,Zagheni:2008,Read:2012,Danon:2013}
or the construction of synthetic
populations from socio-demographic data \cite{Iozzi:2010,Fumanelli:2012},  which have yielded important insights and allowed
to inform large-scale epidemiological models.

On the one hand, surveys are costly and often have often a low response \cite{Danon:2013}, 
and the precise formulation of the question might influence the answers. 
Answers are also subject to memory biases, which are difficult to estimate \cite{Smieszek:2012} but can be large \cite{Smieszek:2014}. 
Moreover, surveys often collect ego-networks on single days (see however
\cite{Read:2008,Smieszek:2012}), and it is then difficult to estimate some properties of the contact 
networks known to be relevant for the spread of infectious diseases, 
such as the number of triangles and the fraction of repeated contacts from one day to the next \cite{Stehle:2011a,Smieszek:2009}.
In contrast, the cost of the deployment of wearable sensors is nowadays affordable, they can be used during several consecutive days
within a specific population,
and they provide an objective definition of contact that does not rely on memory of individuals.

On the other hand, surveys yield large-scale data sets providing age-based contact matrices of crucial importance in parametrizing models
of infectious diseases. Notably, surveys can include a distinction between different types of contact, and in particular 
inform on  the occurrence of physical contacts, which is known to be important in epidemiology
\cite{Melegaro:2012,Goeyvaerts:2010}, while 
wearable sensors do not yield such information. Deployments of wearable sensors are moreover typically limited
to relatively small-scale settings and specific contexts, and do not give information on contacts
occurring between individuals participating to the measures and other individuals (e.g., here, students from
non-participating classes). Another potential issue concerns the possibility that individuals 
changed their behaviour because they were wearing badges and knew they were participating in a scientific measure.
Unfortunately, it is impossible to completely rule out this potential bias
as an accurate check would require monitoring an independent data source for face-to-face
contacts such as video, and because of scalability issues this would be feasible
only for small control groups. The robustness of a series of statistical characteristics across contexts involving different types of participants
\cite{Cattuto:2010,Stehle:2011a,Stehle:2011b,Isella:2011a,Isella:2011b,Vanhems:2013,Salathe:2010}
tends in this respect to be reassuring.

Overall, the use of sensors appears as complementary to the use of surveys, in particular by providing information at a finer scale:
they indicate how far schools, classes or workplaces are from being fully mixed, as the data give access to the density of the contact
networks of each class and between classes; they give 
reliable information with high temporal resolution
on contact durations  and on their statistical distributions, shown to be highly heterogeneous,
an important property in the context of infectious disease transmissions.
In this respect, further data gathering efforts in different contexts are thus certainly of interest, in particular
to further assess the robustness of our results and to compare data obtained in different contexts.
Furthermore,  the combination of such data with other data 
sources such as surveys might represent an interesting further step. 
}

\section*{Acknowledgments}

We are grateful to the SocioPatterns collaboration~\cite{SocioPatterns} for providing access to the SocioPatterns sensing platform that
was used in collecting the contact data, and to C. Cattuto for discussions.
We thank the students of Lyc\'ee Thiers in Marseilles, France, who accepted to participate to the data collection.


\begin{thebibliography}{99}

\bibitem{Longini:1982}
Longini IM Jr, Koopman JS, Monto AS, Fox JP (1982) 
Estimating household and community transmission parameters for influenza.
Am J Epidemiol 115(5): 736-751.

\bibitem{Viboud:2004}
Viboud C, Bo\"elle PY, Cauchemez S, Lavenu A, Valleron AJ, et al. (2004) 
Risk factors of influenza transmission in households. 
Br J Gen Pract 54(506): 684-689.


\bibitem{Read:2012}
Read JM, Edmunds WJ, Riley S, Lessler J, Cummings DAT (2012).
Close encounters of the infectious kind: methods to measure social mixing behaviour. 
Epidemiology and Infection, 140:2117-2130.

\bibitem{Edmunds:1997} 
Edmunds WJ, O'Callaghan CJ, Nokes DJ (1997)
Who mixes with whom? A method to determine the contact patterns of adults that may lead to the spread of airborne infections. 
Proc Biol Sci  264:949-57.

\bibitem{Mossong:2008} 
Mossong J, Hens N, Jit M, Beutels P, Auranen K, et al. (2008)
Social contacts and mixing patterns relevant to the spread  of infectious diseases. 
PLoS Med  5:e74.
   
\bibitem{Read:2008} 
Read JM, Eames KT, Edmunds WJ (2008)
{Dynamic social networks and the implications for the spread of infectious disease}. 
J R Soc Interface  5: 1001-7.

\bibitem{Zagheni:2008} 
Zagheni E, Billari FC, Manfredi P, Melegaro A, Mossong J, et al. (2008)
{Using time-use data to parameterize models for the spread of close-contact infectious diseases}. Am J Epidemiol 
168: 1082-90.

\bibitem{Mikolajczyk:2008} 
Mikolajczyk RT, Akmatov MK, Rastin S, Kretzschmar M (2008) Social 
contacts of school children and the transmission of respiratory-spread pathogens. Epidemiol Infect 136(6): 813-822

\bibitem{Conlan:2011} 
Conlan AJ, Eames KT, Gage JA, von Kirchbach JC, Ross JV, et al. (2011)
Measuring social networks in British primary schools through scientific engagement. Proc R Soc B 278: 1467-75.

\bibitem{Smieszek:2012}
Smieszek T, Burri EU, Scherzinger R, Scholz RW (2012) Collecting close-contact social mixing data with contact diaries: 
reporting errors and biases. Epidemiol Infect 140: 744-752.

\bibitem{Potter:2012} Potter GE, Handcock MS, Longini IM Jr., Halloran ME (2012)
Estimating within-school contact networks to understand influenza transmission. 
Ann Appl Stat.  6(1): 1-26.

\bibitem{Danon:2013}
 Danon L, Read JM, House TA, Vernon MC, Keeling MJ.(2013) 
 Social encounter networks: characterizing Great Britain. 
 Proc R Soc B  280: 20131037.
 
 \bibitem{Smieszek:2014} 
Smieszek T, Barclay V C, Seeni I, Rainey J J, Gao H, et al. (2014) 
How should social mixing be measured? Comparing survey- and sensor-based methods. 
BMC Infectious Diseases 14:136.
 
\bibitem{Pentland:2008}
Pentland A (2008) Honest signals: how they shape our world. Cambridge, MA: MIT Press.

\bibitem{Salathe:2010} 
Salath\'e M, Kazandjieva M, Lee J W, Levis P, Feldman M W, et al. (2010) 
A high-resolution human contact network for infectious disease transmission. 
PNAS  107 (51) 22020-22025. doi: 10.1073/pnas.1009094108

\bibitem{Hornbeck:2012}
Hornbeck T, Naylor D, Segre AM, Thomas G, Herman T, et al.. (2012) 
Using Sensor Networks to Study the Effect of Peripatetic Healthcare Workers on the Spread of Hospital-Associated Infections.
J Infect Dis 206:1549.

\bibitem{Sekara:2014}
Stopczynski A, Sekara V, Sapiezynski P, Cuttone A, Madsen MM, et al. (2014)
Measuring Large-Scale Social Networks with High Resolution. PLoS ONE 9(4): e95978. 


\bibitem{Barclay:2014}
Barclay VC, Smieszek T, He J, Cao G, Rainey JJ, et al. (2014) 
Positive Network Assortativity of Influenza Vaccination at a High School: Implications for Outbreak Risk and Herd Immunity. 
PLoS ONE 9(2): e87042. 

\bibitem{SocioPatterns} SocioPatterns website. Available: \verb+http://www.sociopatterns.org/+. Accessed 2014 Aug. 26.

\bibitem{Cattuto:2010} Cattuto C, Van den Broeck W, Barrat A, Colizza V, Pinton JF, et al. (2010) 
Dynamics of person-to-person interactions from distributed RFID sensor networks. 
PLoS One 5: e11596. doi: 10.1371/journal.pone.0011596

\bibitem{Stehle:2011a} Stehl\'e J, Voirin N, Barrat A, Cattuto C, Colizza V, et al. (2011) 
Simulation of an SEIR infectious disease model on the dynamic contact network of conference attendees. 
BMC Med 9: 87. 
doi: 10.1186/1741-7015-9-87

\bibitem{Stehle:2011b} Stehl\'e J, Voirin N, Barrat A, Cattuto C, Isella L, et al. (2011) 
High-resolution measurements of face-to-face contact patterns in a primary school. 
PLoS One 6: e23176. 

\bibitem{Isella:2011a} Isella L, Romano M, Barrat A, Cattuto C, Colizza V, et al. (2011) 
Close encounters in a pediatric ward: measuring face-to-face proximity and mixing patterns with wearable sensors. 
PLoS One 6: e17144. 

\bibitem{Isella:2011b} 
Isella L, Stehle J, Barrat A, Cattuto C, Pinton JF, et al. (2011) 
What's in a crowd? Analysis of face-to-face behavioral networks. 
J. Theor. Biol. 271: 166-180.

\bibitem{Vanhems:2013} 
Vanhems P, Barrat A, Cattuto C, Pinton JF, Khanafer, et al. (2013) 
Estimating potential infection transmission routes in hospital wards using wearable proximity sensors. 
PLoS One 6: e73970. doi: 10.1371/journal.pone.0073970

\bibitem{Stehle:2010}
Stehl\'e J, Barrat A, Bianconi G (2010)
Dynamical and bursty interactions in social networks.
Phys. Rev. E  81, 035101(R).

\bibitem{Zhao:2011}
Zhao K, Stehl\'e J, Bianconi G, Barrat A (2011)
Social network dynamics of face-to-face interactions.
Phys. Rev. E 83, 056109.

\bibitem{Starnini:2013}
Starnini M, Baronchelli A, Pastor-Satorras R (2013)
Modeling Human Dynamics of Face-to-Face Interaction Networks.
Phys. Rev. Lett. 110:168701.

\bibitem{Barabasi:2005}
Barab\`asi A-L (2005) 
The origin of bursts and heavy tails in human dynamics, Nature 435 (7039), 207.

\bibitem{Barabasi:2010}
Barab\'asi A-L (2010)
Bursts: The Hidden Pattern Behind Everything We Do.  Dutton Adult.

\bibitem{Iozzi:2010}
Iozzi F, Trusiano F, Chinazzi M, Billari FC, Zagheni E, et al. (2010) 
Little Italy: An Agent-Based Approach to the Estimation of Contact Patterns- Fitting Predicted Matrices to Serological Data. 
PLoS Comput Biol 6(12): e1001021. 

\bibitem{McPherson:2001} 
McPherson M, Smith-Lovin L, Cook JM (2001)
Birds of a Feather: Homophily in Social Networks. 
Annual Review of Sociology  27: 415-444.

\bibitem{Stehle:2013} Stehl\'e J, Charbonnier F, Picard T, Cattuto C, Barrat A. (2013) 
Gender homophily from spatial behavior in a primary school: a sociometric study. 
Soc. Net. 35: 604.

\bibitem{Blower:2011}
Blower S, Go MH (2011) 
The importance of including dynamic social networks when 
modeling epidemics of airborne infections: does increasing complexity increase accuracy? BMC Med 9: 88.

\bibitem{Machens:2013}
Machens A, Gesualdo F, Rizzo C, Tozzi AE, Barrat A, et al. (2013) 
An infectious disease model on empirical networks of human contact: bridging the gap between dynamic network data and contact matrices. 
BMC Infect Dis 13: 185.


\bibitem{Maslov:2004}
Maslov, S., Sneppen, K., Zaliznyak, A. (2004) Detection of topological patterns
  in complex networks: correlation profile of the {I}nternet. 
  Physica A 333: 529--540.
  
\bibitem{Gemmetto}
Gemmetto V, Barrat A, Cattuto C, in preparation.

\bibitem{Fumanelli:2012}
Fumanelli L, Ajelli M, Manfredi P, Vespignani A, Merler S (2012) 
Inferring the Structure of Social Contacts from Demographic Data in the Analysis of Infectious Diseases Spread. 
PLoS Comput Biol 8(9): e1002673. 

\bibitem{Smieszek:2009}
Smieszek T, Fiebig L, Scholz RW (2009) 
Models of epidemics: when contact repetition and clustering should be included. 
Theor Biol Med Model  6: 11. 

\bibitem{Melegaro:2012}
Melegaro A, Juta M, Gaya N, Zagheni E, Edmunds WJ (2011)
What types of contacts are important for the spread of infections? Using contact survey data to explore European mixing patterns.
Epidemics 3: 143-151.
    
\bibitem{Goeyvaerts:2010}
Goeyvaerts N, Hens N, Ogunjimi B, Aerts M, Shkedy Z, et al. (2010)
Estimating infectious disease parameters from data on social contacts and serological status.
J. Roy. Stat. Soc. 59: 255-277.


\end{thebibliography}
\end{document}